\documentclass[fleqn,usenatbib]{mnras}

\usepackage{newtxtext,newtxmath}

\usepackage[T1]{fontenc}
\usepackage{ae,aecompl}

\usepackage{dsfont}
\usepackage{pdflscape}

\usepackage{graphicx}	
\usepackage{amsmath}	
\usepackage{amssymb}	

\newcommand {\hi} {\ion{H}{i}}
\newcommand {\hii} {\ion{H}{ii}}
\newcommand {\hei} {\ion{He}{i}}
\newcommand {\heii} {\ion{He}{ii}}
\newcommand {\heiii} {\ion{He}{iii}}
\newcommand {\htwo}{H$_2$} 

\newcommand {\eqcnc} {\,\mathrm{cm}^{-3}}
\newcommand {\Pt} {{\bf{\mathds{P}}}}

\title[Molecular chemistry and radiation hydrodynamics]{A simple model for molecular hydrogen chemistry coupled to radiation hydrodynamics}

\author[Sarah Nickerson et al.]{
Sarah Nickerson,$^{1}$\thanks{E-mail: snickers@physik.uzh.ch}
Romain Teyssier,$^{1}$
Joakim Rosdahl$^{2,3}$
\\
$^{1}$Institute for Computational Science, University of Z{\"u}rich, Winterthurerstrasse 190, CH-8057 Z{\"u}rich, Switzerland\\
$^{2}$Leiden Observatory, Leiden University, PO Box 9513, NL-2300 RA, Leiden, the Netherlands\\
$^{3}$Centre for Astronomy Research, University of Lyon, 9 avenue Charles Andr{\'e}, 69230 Saint-Genis-Laval, France\\
}

\date{Accepted XXX. Received YYY; in original form ZZZ}

\pubyear{2018}

\begin{document}
\label{firstpage}
\pagerange{\pageref{firstpage}--\pageref{lastpage}}
\maketitle

\begin{abstract}
We introduce non-equilibrium molecular hydrogen chemistry into the radiation hydrodynamics code \textsc{Ramses-RT}. This is an adaptive mesh refinement grid code with radiation hydrodynamics that couples the thermal chemistry of hydrogen and helium to moment-based radiative transfer with the Eddington tensor closure model. The \htwo\ physics that we include are formation on dust grains, gas phase formation, formation by three-body collisions, collisional destruction, photodissociation, photoionization, cosmic ray ionization, and self-shielding. In particular, we implement the first model for \htwo\ self-shielding that is tied locally to moment-based radiative transfer by enhancing photodestruction. This self-shielding from Lyman-Werner line overlap is critical to \htwo\ formation and gas cooling. We can now track the non-equilibrium evolution of molecular, atomic, and ionized hydrogen species with their corresponding dissociating and ionizing photon groups. Over a series of tests we show that our model works well compared to specialized photodissociation region codes. We successfully reproduce the transition depth between molecular and atomic hydrogen, molecular cooling of the gas, and a realistic Str{\"o}mgren sphere embedded in a molecular medium. In this paper we focus on test cases to demonstrate the validity of our model on small scales. Our ultimate goal is to implement this in large-scale galactic simulations.
\end{abstract}




\section{Introduction}
\label{sec:intro}

The study of \htwo\ in galaxies touches on an immense range of scales. Observations on the galactic kpc scale show that \htwo\ correlates with star formation \citep{Wong2002, Schruba2011, Leroy2013}. On the pc scale are the molecular clouds themselves. Within the Milky Way their mass distribution follows a power law similar to that of the luminosity distribution of OB stars \citep{Williams1997}, and their velocity dispersion follows a power law that increases with radius \citep{Larson1981}. These intermediate scale mechanics are influenced by, and in turn influence, both the galaxy-wide dynamics and the molecular-level chemistry.

The typical giant molecular cloud (GMC) has an outer layer of atomic hydrogen (\hi), an inner core of \htwo, and a deeper CO core. GMCs have long been established to be the sites of star formation  \citep{McKee2007}. \citet{Schmidt1959} and \citet{Kennicutt1998} demonstrate that the  neutral hydrogen surface density correlates to the surface density of star formation (the K-S relation), while recent observations show that the \htwo\ surface density correlates even more tightly to star formation \citep[e.g.][]{Bigiel2008}. 

It is unclear whether there is a causation behind this correlation. What is known, however, is that \htwo\ is an important coolant for interstellar gas \citep{Gnedin2011}. Unfortunately, \htwo\ is exceptionally difficult to directly observe owing to its lack of a dipole moment. This leads to GMC identification by CO content \citep{Solomon1987}. The \htwo\ content can then be inferred by a conversion factor between the CO intensity and the column density of \htwo. This conversion factor has been extensively measured and is found to be constant for the Milky Way. However, further studies beyond the Milky Way show that it might depend on galaxy morphology and metallicity \citep{Bolatto2013}.

Smaller still than GMCs is the scale of the particles themselves and the chemistry by which hydrogen becomes molecular. Grains of interstellar dust serve as catalysts by which \hi\ sticks and coalesces into \htwo\ (\citealt{Gould1963}; see \citealt{Wakelam2017} for a comprehensive review). \htwo\ may also form by gas phase interactions, but this process is much slower and was only important in the early Universe when metals were scarce \citep{Galli1998}. At high temperatures ($T \gtrsim 1000$ K), collisions between \htwo\ and other particles dissociate \htwo\ into \hi\ \citep{Glover2008}. \htwo\ is also ionized by high-energy photons, $h\nu \geq 15.42$ eV, where $\nu$ is the photon frequency \citep{Abel1997}. Finally, photons that fall into the Lyman-Werner (LW) band, 11.2 to 13.6 eV, dissociate \htwo. Outer regions of the GMCs absorb the LW photons at stronger wavelengths first, forming an \hi\ layer, but allow weaker wavelengths to penetrate further until they too are absorbed.

In addition to photodestruction by absorption, two processes shield \htwo\ from radiation. The first is shielding by dust, and the second is \htwo\ self-shielding. Only about 10 per cent of the LW absorption leads to \htwo\ dissociation \citep{Stecher1967} and the rest of the photons are destroyed without contributing to photodissociation \citep{Sternberg2014}. The absorption rate is highly dependent on the wavelength of the LW band \citep{Abgrall1992,Haiman2000}. Certain bands become optically thick at high \htwo\ column densities, and dissociation is quashed, while bands with weaker absorption can still penetrate the cloud. Here an increase in the natural line width, due to Heisenberg uncertainty, leads to interference between the LW bands. Self-shielding is weakest at low column densities and increases further into the cloud. Hence, \htwo\ self-shielding functions calculated from experiments are given in terms of the column density of \htwo, the most widely used of which is from \citet{Draine1996}. \citet{Gnedin2014a} update this function to account for turbulence in molecular clouds.

It is this range of scales, from the quantum mechanical nature of \htwo\ self-shielding, to the far-reaching gravitational influence of a galaxy on GMCs, that makes simulating molecular chemistry challenging. It is modelled on the smallest scale in photodissociation region (PDR) codes and on the largest scale in galaxy codes.

PDRs are predominately neutral regions of the interstellar medium (ISM) in which far-ultraviolet (UV) photons (6 $< h\nu <$ 13.6 eV) control the temperature and chemistry. They contain all of the atomic and at least 90 per cent of the molecular gas in the Milky Way, and are a major non-stellar source of infrared (IR) radiation in the ISM \citep{Hollenbach1999}. PDR models are diverse, exhibiting different geometries, from one-dimensional to spherical, and are developed to study a range of phenomena. Many focus on interstellar clouds: both the clumps inside the clouds themselves and the boundaries between molecular clouds and ionized regions. Others study plasmas, circumstellar discs, planetary nebulae, the centre of the Milky Way, and the ratio between CO and \htwo. PDR models involve sophisticated chemical networks with species of hydrogen, carbon, oxygen, and silicone; detailed treatment of dust; radiative transfer (RT) of multiple photon groups; and the heating and cooling processes a cloud undergoes as a result of the interactions between the gas, dust, and photons. \citet{Roellig2007} is a first of its kind comparison study of 10 PDR codes, consisting of a series of benchmark tests to highlight where the codes converge and to understand why they differ.

The past decade has seen a number of methods to model the \htwo\ chemistry in both semi-analytical galaxy models and hydrodynamical galaxy simulations. They explore the nature of the relationship between star formation and \htwo, test star formation recipes, and see how \htwo\ affects the gas composition of galaxies. The semi-analytical models use equilibrium equations to find the \htwo\ fraction \citep{Fu2010, Somerville2015,Xie2017}  while hydrodynamic simulations use either equilibrium equations \citep{Pelupessy2006,Robertson2008,Kuhlen2012,Halle2013,Hopkins2014,Thompson2014} or a series of non-equilibrium chemical networks \citep{Gnedin2009,Gnedin2011,Christensen2012,Tomassetti2014,Baczynski2015,Richings2016, Hu2016,Capelo2018,Lupi2018,Pallottini2017,Katz2017}. Equilibrium calculations have the advantage of speed but use the assumption that the chemical species are in equilibrium with their environment. Non-equilibrium codes instead use local rates of destruction and creation of chemical species, and networks of rate equations.

To date, only four of these codes use radiative transfer with a non-equilibrium chemical network. \citet{Gnedin2011} implement an \htwo-based star formation recipe for cosmological galaxy simulations and demonstrate that the molecular content of a galaxy and its K-S relation are sensitive to both the dust to gas ratio and the UV flux. In contrast \citeauthor{Lupi2018}'s (\citeyear{Lupi2018}) star formation recipe is independent of \htwo\ content and still reproduces the K-S relation. Both these use a moment-based method for radiative transfer. \citet{Baczynski2015} instead use ray tracing, and provide the non-equilibrium chemistry of hydrogen and carbon. \citeauthor{Katz2017}'s (\citeyear{Katz2017}) method is the most similar to ours, modifying \textsc{Ramses-RT} to track \htwo\ in cosmological simulations for comparison to observations with the Atacama Large Millimeter/submillimeter Array. An earlier \htwo\ implementation in \textsc{Ramses} \citep{Valdivia2015} looks at only galaxy segments with radiation from an external UV field. Our \htwo\ model differs from all of these in the \htwo\ self-shielding approximation.

The models for \htwo\ self-shielding, as mentioned above, describe the shielding as a function of cloud column density. Simulations, however, use the volume density, and conversion is necessary. The most computationally simple way is to convert the volume density into a column density using a Jeans length, a Sobolev length, or a Sobolev-like length. Non-local methods use neighbouring cells to compute a column density, but are more expensive (\citet{Wolcott-Green2011} provide an overview). \citet{Gnedin2011} treat their conversion length as a free parameter to be computed, while \citet{Lupi2018} use a Jeans length, and \citet{Katz2017} use the cell width. Only \citet{Baczynski2015} avoid the need for volume to column conversion because they use ray tracing to compute the \htwo\ column density directly. Nonetheless, this radiative transfer method is computationally expensive in large simulations with multiple sources.  These self-shielding functions all decrease the \htwo\ destruction.

In this work, we present the first model of \htwo\ physics tied directly to moment-based radiative transfer by a local self-shielding approximation to enhance photodestruction in the LW band. In this way we do not need to use a volume-to-column density approximation as previous codes have. \textsc{Ramses-RT} is optimized for radiation-hydrodynamical galaxy simulations and photoionization, but our new method also holds up under the conditions of PDR codes, thus linking the two regimes. Both radiative transfer (RT) and non-equilibrium calculations of \htwo\ are important to study the problem of how \htwo\ affects galaxies. We will be able to use this methodology for not only isolated disc galaxies, but also galaxies in a cosmological context. The combination of moment-based RT and the few photon groups required uniquely situates us to simulate \htwo\ chemistry in cosmology.

\textsc{Ramses} \citep{Teyssier2002} is an adaptive mesh refinement (AMR)  code for $N$-body hydrodynamical galaxy simulations, both cosmological and isolated discs. \textsc{Ramses-RT} \citep{Rosdahl2013} implements radiation hydrodynamics for \textsc{ramses}, coupling photons to the non-equilibrium chemistry of the neutral and ionized species of hydrogen and helium. It utilizes a moment-based method of radiative transfer, which unlike ray tracing is independent of the number of sources. This makes it ideal for galaxy simulations that host large numbers of stars. In this paper, the first of two, we  present an upgrade to implement \htwo\ chemistry in the \textsc{Ramses-RT} code. Our tests  show our \htwo\ model's ability to match PDR code benchmarks and simulate realistic molecular Str{\"o}mgren spheres. In a follow-up paper we will demonstrate the effects of our \htwo\ model in galaxy simulations.

In Section \ref{sec:meth}, we give an overview of \textsc{Ramses-RT} and our new implementation for \htwo\ physics. In Section \ref{sec:tests}, idealized tests prove the rigour of our method. We include comparisons to PDR codes and a Str{\"o}mgren sphere embedded in a molecular medium. Finally, in Section \ref{sec:conc}, we summarize our findings and provide future directions for our current work.

\section{Method}
\label{sec:meth}

We begin with an overview of \textsc{ramses} and its radiative transfer features before diving into the specific details of the \htwo\ physics. Previously, \textsc{Ramses-RT} only tracked \hi, \hii, \hei,  \heii, and \heiii. In this review we show where \htwo\ is also included in the equations in order to provide a complete and updated picture.

\subsection{Overview of \textsc{Ramses-RT}}
\textsc{Ramses} is an adaptive mesh refinement (AMR) hydrodynamical code with an $N$-body solver for stellar populations and dark matter, and a tree-based data structure grid for the gravitational potential and advection of gas \citep{Teyssier2002}.  The radiative transfer extension \citep{Rosdahl2013} introduces radiative transfer coupled to the hydrodynamics in \textsc{Ramses}, directly tracking photon groups that are tied to the non-equilibrium chemistry of \hi, \hii, \hei,  \heii, and \heiii\ via photoionization and heating. In this paper we introduce the non-equilibrium chemistry of \htwo\ and include its index in this overview section. The full details of the \htwo\ chemistry are in Section \ref{sec:molrecipe}. 

In \textsc{Ramses-RT}, radiation frequency is discretized into groups whose attributes are averaged over frequency ranges. Each gas cell at a given time is described by a state $ \mathcal{U}=(\rho, \rho\textbf{u}, E, \rho x_{\text{\hi}}, \rho x_{\text{\hii}}, \allowbreak \rho x_{\text{\heii}}, \rho x_{\text{\heiii}}, N_i, \textbf{F}_i)$ (mass density, momentum density, energy density, \hi\ fraction abundance, \hii\ fraction abundance, \heii\ fraction abundance,  \heiii\  fraction abundance, photon density for each radiation  group, and flux for each group). \htwo\ and \hei\ fractions are not tracked, but can be recovered from the fractions of other species. 

\textsc{Ramses-RT} uses a moment-based approach by treating the photons as a fluid, which renders the computational cost independent of the number of radiation sources. For galaxy simulations filled with stars, this makes a moment-based method much faster compared to the alternative of ray tracing. The main disadvantage of the moment-based method is that we need an approximate closure model for the pressure tensor (equation \ref{eqn:rtf}). An exact treatment requires ray tracing, which is computationally expensive, and we opt instead for a local method. One further hurdle is that in \textsc{Ramses-RT} the radiative transfer is advanced explicitly in time, and in the free-streaming limit this leads to much smaller time-steps for the RT as compared to pure hydrodynamic simulations. \textsc{Ramses-RT} solves this problem with the reduced speed of light approach \citep{Gnedin2001}, which is a valid approximation as long as the light crossing time is shorter than the sound crossing, recombination, and advection time-scales. 

\textsc{Ramses-RT} implements recombination emission from every gas cell, and it also provides the option of an on-the-spot approximation (OTSA) where recombination photons are assumed to be absorbed in the same gas cell, thereby ignoring direct-to-ground-state recombinations. A later extension \citep{Rosdahl2015a} adds radiation pressure and dust absorption to \textsc{Ramses-RT}.

\subsection{Moment-based radiative transfer}
We summarize here the moment-based approach in \textsc{Ramses-rt} as described in \citet{Rosdahl2013}, with the addition of molecular hydrogen. Further sources, \citet{Mihalas1984} and \citet{Aubert2008} outline this process in more detail.  

$I_{\nu}(\textbf{x},\textbf{n} , t)$ is the specific radiation intensity at a wavelength $\nu$, location $\textbf{x}$, direction $\textbf{n}$, and time $t$ in units erg cm$^{-2}$ s$^{-1}$ Hz$^{-1}$ rad$^{-2}$. The evolution of the specific intensity is described by the equation of radiative transfer:

\begin{equation}\label{eqn:rttrans}
\frac{1}{c_{\text{r}}} \frac{\partial I_{\nu}}{\partial t}
+ \textbf{n} \cdot \nabla I_{\nu}
= -\kappa_{\nu}I_{\nu} + \eta_{\nu},
\end{equation}
where $c_{\text{r}}$ is the reduced speed of light, $\kappa_{\nu}(\textbf{x},\textbf{n},t)$ is the gas opacity, and $\eta_{\nu}(\textbf{x},\textbf{n},t)$ is the source function. The time evolution of the photon number density $N_{\nu}$ and flux $\textbf{F}_{\nu}$ are then extracted from equation \eqref{eqn:rttrans} by taking the zeroth and first angular moments:

\begin{align}
\frac{\partial N_{\nu}}{\partial t}+\nabla \cdot \textbf{F}_{\nu} &= -\sum_j^{\text{\htwo,\hi,\hei,\heii}} {n_j \sigma_{\nu j}c_{\text{r}} N_{\nu}}\label{eqn:rtn}\\
& - \kappa_{\text{P}} \rho_{\text{d}} c_{\text{r}} N_{\nu} + \dot{N}^{\star}_{\nu} + \dot{N}^{rec}_{\nu}, \nonumber \\
\frac{\partial \textbf{F}_{\nu}}{\partial t} +{c_{\text{r}}^2}\nabla \cdot \Pt_{\nu} &= -\sum_j^{\text{\htwo,\hi,\hei,\heii}}{n_j \sigma_{\nu j}c_{\text{r}} \textbf{F}_{\nu}} \label{eqn:rtf} \\
&-\kappa_{\text{R}} \rho_{\text{d}} c_{\text{r}} \textbf{F}_{\nu} \nonumber,
\end{align}
where $n_j$ is the number density of species $j$, $\sigma_{\nu j}$ is the ionization/dissociation cross-section between photons with frequency $\nu$ and species $j$, $\kappa_{\text{P}}$ and $\kappa_{\text{R}}$ are the Planck and Rosseland dust opacities,  $\rho_{\text{d}}$ is the dust volume density, $\dot{N}^{\star}_{\nu}$ is the number of photons injected by stars, $\dot{N}^{rec}_{\nu}$ is the number of photons injected by gas recombination when OTSA is off, and $\Pt_{\nu}$ is the radiative pressure tensor. The dust density is given by $ \rho_{\text{d}} \equiv Z f_{\text{d}} \rho$, where $Z$ is the metallicity, $f_{\text{d}}$ is the fraction of gas that holds dust, and $\rho$ is the gas volume density. We currently do not track dust independently and use the ionization state of the gas as an indication of the dust content,
\begin{equation}
f_{\text{d}}=1-x_{\text{\hii}}. \label{eqn:fdust}
\end{equation}

Equations \eqref{eqn:rtn} and \eqref{eqn:rtf} are continuous in $\nu$, but for the purposes of computation we deal with photon groups whose properties are averaged over their entire range. We replace $N_{\nu}$ and $F_{\nu}$ with $N_i$ and $F_i$, which are the integrated sums over the range. The choice of photon groups for \textsc{Ramses-RT} is easily customized. Mainly we are concerned with four groups: (1), the LW band of \htwo-dissociating radiation (11.2 eV to 13.60 eV); (2), \hi-ionizing (13.60 eV to 24.59 eV); (3), \hei-ionizing (24.59 eV to 54.42 eV); and (4), \heii-ionizing (54.42 eV and above) radiation. \hi\ and \htwo\ are ionized by groups 2, 3, and 4; \hei\ by groups 3 and 4; and \heii\ by group 4.

The pressure tensor, $\Pt_{\nu}$, closes equations \eqref{eqn:rtn} and \eqref{eqn:rtf} and is usually the product of the photon number density and the Eddington tensor, for which several approximations exist. We use the M1 closure relation \citep{Levermore1984}, further details of which are given in \citet{Rosdahl2013}.

Equations \eqref{eqn:rtn} and \eqref{eqn:rtf} are solved for each time-step and photon group by an operator-splitting strategy. The photon flux and density and species abundances are updated in a fixed order: photon injection, photon transport, and thermochemistry.

The \textbf{photon injection} step solves a single equation,
\begin{equation}\label{eqn:inj}
\frac{\partial N_i}{\partial t}=\dot{N}^{\star}_i,
\end{equation}
to account for all the photons injected into a cell, usually in galaxy simulations by stellar sources. This is carried out discretely over each photon group $i$, and sums over all the photon sources in the cell.

In the \textbf{transport step} the photons are treated as free-flowing between cells, described by the equations,
\begin{align}
\frac{\partial N_i}{\partial t}+\nabla \cdot \textbf{F}_i &= 0, \label{eqn:ntransport} \\
\frac{\partial \textbf{F}_i}{\partial t}+c_{\text{r}}^2\nabla \cdot \Pt_i&=0.\label{eqn:ftransport}
\end{align}
There are many functions available to solve these equations for the intercell flux. \textsc{Ramses-RT} provides two options. The Harten-Lax-van Leer (HLL) flux function \citep{Harten1983} is ideal for modelling beams and shadows, but shows asymmetries for isotropic sources. The global Lax-Friedrich (GLF) function \citep{Lax1954} is better suited for isotropic sources and preserves symmetry, but tends to diffuse beams. Both these functions are useful depending on the scenario.

Finally, the \textbf{thermochemistry step} handles the interactions between the photons, gas temperature, dust, and the gas species \htwo, \hi, \hii, \hei, \heii, and \heiii. Equations \eqref{eqn:rtn} and \eqref{eqn:rtf}  are solved without the injection or divergence terms. Non-equilibrium chemistry equations are too stiff to solve explicitly, due to time-scales differing by orders of magnitude which render the time-steps too small. Instead, we solve the chemistry semi-implicitly in a specific order based on the algorithm of \citet{Anninos1997}. This algorithm involves a backward differencing scheme for solving the collisional and radiative processes of hydrogen and helium species. It considerably speeds up computation compared to a packaged solver, while not sacrificing accuracy. \textsc{Zeus-MP} \citep{Whalen2006} and \textsc{RH1D} \citep{Ahn2008} both use the \citet{Anninos1997} method and compare well with other radiative transfer codes \citep{Iliev2006,Iliev2009}. We will expand on this step in Section \ref{sec:thermostep}, after first introducing the details of \htwo\ chemistry in Section \ref{sec:molrecipe}.

\subsection{Molecular hydrogen recipe}
\label{sec:molrecipe}
In this section we describe in detail the new \htwo\ chemistry implemented into \textsc{Ramses-RT}. For the three major species of hydrogen the reaction rates are given by
\begin{align} 
\dot{n}_{\text{\hii}}&=-\dot{n}_{\text{\hi}}-2\dot{n}_{\text{\htwo}}, \label{eqn:hiirate}\\
\dot{n}_{\text{\hi}}&=\alpha_{\text{\hi}}(T) n_{\text{e}}n_{\text{\hii}}-\beta_{\text{\hi,e}}(T)n_{\text{e}}n_{\text{\hi}} \label{eqn:hirate} \\
&-\Gamma_{\text{\hi}}(N_{\text{\hi}})n_{\text{\hi}}-\xi_{\text{\hi}}n_{\text{\hi}}-2\dot{n}_{\text{\htwo}}, \nonumber \\ 
\dot{n}_{\text{\htwo}}&=\alpha_{\text{\htwo}}^{\text{Z}}(T) Z f_{\text{d}} n_{\text{H}}n_{\text{\hi}}+ \alpha_{\text{\htwo}}^{\text{GP}}(T)n_{\text{\hi}}n_{\text{e}} \label{eqn:h2rate} \\
&+ \beta_{\text{3B}}(T) n_{\text{\hi}}^2(n_{\text{\hi}} + n_{\text{\htwo}}/8) \nonumber \\
&- \beta_{\text{\htwo\hi}}(T)n_{\text{\hi}}n_{\text{\htwo}} -\beta_{\text{\htwo\htwo}}(T)n_{\text{\htwo}}^2 \nonumber \\
&-\Gamma_{\text{\htwo}}^{\text{LW}}(N_{\text{\htwo}})n_{\text{\htwo}}-\Gamma_{\text{\htwo}}^+(N_{\text{\hi}})n_{\text{\htwo}}- \xi_{\text{\htwo}}n_{\text{\htwo}}\nonumber,
\end{align}
where $n$ is the number density of the subscript species (\htwo, \hi, \hii, or $e$ for electrons), $\alpha$ is the formation/recombination rate of the subscript species, $\beta$ the collisional dissociation/ionization rate between the two subscript species, $\beta_{\text{3B}}$ is the collisional rate for three-body interactions, $\Gamma$ is the photoionization/dissociation rate of the subscript species, $N$ is the number density of the photodissociating/ionizing photon group(s) of the subscript species, $\xi$ is the cosmic ray ionization rate for the subscript species, $T$ is the temperature, $Z$ is the metallicity as a fraction of solar, and $f_{\text{d}}$ is the dust fraction (equation \ref{eqn:fdust}).

\htwo\ requires two creation terms: $\alpha_{\text{\htwo}}^{\text{Z}}(T)$ for formation on dust grains and $\alpha_{\text{\htwo}}^{\text{GP}}(T)$ for gas phase formation. It also has two separate photodestruction terms: $\Gamma_{\text{\htwo}}^{\text{LW}}(N_{\text{\htwo}})$ for dissociation by LW photons, and $\Gamma_{\text{\htwo}}^+(N_{\text{\hi}})$ for photoionization by the same photon groups that ionize \hi. However, as we will explain in Section \ref{sec:ratecoeff}, we treat \htwo\ ionization as a dissociation that produces \hi\ and not \hii. 

The rates for formation, collisional ionization, and photoionization of \hi\ are preserved from \citet{Rosdahl2013}. The \htwo\ rate coefficients for formation, collisions, and photodestruction are described in the following section.

\subsubsection{Molecular hydrogen rate coefficients}
\label{sec:ratecoeff}

We draw our \htwo\ coefficients for the rate equations from a wide range of sources:
\begin{align}
\alpha_{\text{\htwo}}^{\text{Z}}(T)&=\frac{9.0\times10^{-17}T_2^{0.5}}{1+0.4T_2^{0.5}+0.2T_2+0.08T_2^2} \text{ cm$^3$s$^{-1}$},\\
\alpha_{\text{\htwo}}^{\text{GP}}(T)&= 8.0\times 10^{-19}T_3^{0.88} \text{ cm$^3$s$^{-1}$}, \\
\beta_{\text{\htwo\hi}}(T)&= 7.073\times 10^{-19} T_{\text{K}}^{2.012}\\
&\times \frac{e^{-5.179\times10^{4}/T_{\text{K}}}}{(1+2.130\times 10^{-5}T_{\text{K}})^{3.512}} \text{ cm$^3$s$^{-1}$},\nonumber\\
\beta_{\text{\htwo\htwo}}(T) &= 5.996\times 10^{-30} T_{\text{K}}^{4.1881}\\
&\times \frac{e^{-5.466\times 10^{4}/T_{\text{K}}}}{(1+6.761\times 10^{-6}T_{\text{K}})^{5.6881}} \text{ cm$^3$s$^{-1}$},\nonumber\\
\beta_{\text{\text{3B}}}(T) &= 6 \times 10^{-32} T_{\text{K}}^{-0.25}+2\times 10^{-31}T_{\text{K}}^{-0.5} \text{ cm$^6$s$^{-1}$}\\
\Gamma_{\text{\htwo}}^{\text{LW}}(N_{\text{\htwo}})&=\sigma^{N}_{1\text{\htwo}}c_{\text{r}}N_{1}  \label{eqn:h2destrate},\\
\Gamma_{\text{\htwo}}^+(N_{\text{\hi}})&=\sum^M_{i=2}\sigma^{N}_{i\text{\htwo}}c_{\text{r}}N_i \label{eqn:h2ionrate}\\
\xi_{\text{\htwo}} &= 7.525 \times 10^{-16} \text{ s}^{-1}, \label{eqn:crh2}\\
\xi_{\text{\hi}} &= 4.45 \times 10^{-16} \text{ s}^{-1},\label{eqn:crhi}
\end{align}
where $T_{\text{K}}=\frac{T}{1K}$, $T_2=\frac{T}{100\text{K}}$, and $T_3=\frac{T}{1000\text{K}}$. $\sigma^{N}_{i\text{\htwo}}$ is the average destruction cross-section between species $i$ and \htwo, $M$ is the total number of photon groups (4 in this paper), and the subscript $1$ refers to the first photon group, which is the LW band in this paper. 

The \htwo\ formation rate as catalysed by dust grains ($\alpha_{\text{\htwo}}^{\text{Z}}$) differs depending on the environment \citep{Wakelam2017}, with the traditional rate $3\times10^{-17}$cm$^3$s$^{-1}$ for diffuse clouds \citep{Jura1974, Gry2002} being lower than the recently measured rate for dense PDRs, $1.5\times10^{-16}$cm$^3$s$^{-1}$ \citep{Habart2004}. In order to encompass all environments, we use the average between these two rates with the functional temperature dependence of \citet{Hollenbach1979}. Gas phase \htwo\ formation is important for low- to zero-metallicity environments ($\alpha_{\text{\htwo}}^{\text{GP}}$) and we use the rate of \citet{McKee2010}, which assumes equilibrium in H$^-$. 

We take into account the collisional destruction between \htwo\ and \hi\ ($\beta_{\text{\htwo\hi}}$) \citep{Dove1986} and itself  ($\beta_{\text{\htwo\htwo}}$) \citep{Martin1998}. The rate for three-body collisions ($\beta_{\text{3B}}$) \citep{Forrey2013} encompasses two processes:
\begin{align}
3\text{\hi} &\rightarrow \text{\hi} + \text{\htwo}, \label{eqn:h3b1}\\
2\text{\hi} + \text{\htwo} &\rightarrow 2\text{\htwo}. \label{eqn:h3b2}
\end{align}
\citet{Forrey2013} gives the rate for process \eqref{eqn:h3b1}, and for process \eqref{eqn:h3b2} \citet{Palla1983} recommend $\beta_{\text{3B}}/8$. Like gas phase \htwo\ formation, three-body collisions have little impact in high-metallicity environments but are relevant for \htwo\ formation in the early Universe with little to no metallicity. We neglect \htwo\ collisions with \hii\ and electrons because \htwo\ is unlikely to coexist with these species.

For the photodissociation of \htwo\ by Lyman-Werner photons ($\Gamma_{\text{\htwo}}^{\text{LW}}$) we use a cross-section derived from the photodissociation rate in \citet{Sternberg2014} and treat it as constant due to the the small range of the LW band,
\begin{equation}
\sigma^{N}_{1\text{\htwo}}=2.1\times 10^{-19} \text{cm$^2$}.
\label{eqn:h2diss_cs}
\end{equation}

The ionization of \htwo\ ($\Gamma_{\text{\htwo}}^+$) occurs via a two step process \citep{Abel1997}. First,
\begin{equation}
\text{\htwo}+\gamma \rightarrow \text{\htwo}^++\text{e}^-,
\label{eqn:h2rx}
\end{equation}
and second one of two processes occurs depending on the frequency of the incident photon:
\begin{align}
\text{\htwo}^+ +\gamma &\rightarrow \text{\hi}+\text{\hii}, \\
\text{\htwo}^+ +\gamma &\rightarrow 2\text{\hii}+\text{e}^-.
\end{align}
However, to depict this entire chain of reactions realistically we would need to track the intermediate species \htwo$^+$ at an added computational cost. In order to keep our methodology simple, we model \htwo\ ionization as the following process:
\begin{align}
\text{\htwo}+\gamma &\rightarrow 2\text{\hi},\label{eqn:h2ion} \\
\text{\hi}+\gamma &\rightarrow \text{\hii}+\text{e}^-. 
\end{align}
Essentially we treat the ionization of \htwo\ as a dissociation. Because the wavelengths that ionize \htwo\ and \hi\ are virtually identical ($h\nu \ge 15.42$ eV for \htwo\ and $h\nu \ge 13.60$ eV for \hi) we assume that the \hi\ produced from ionization of \htwo\ is quickly ionized into \hii. We take the ionization cross-section from \citet{Abel1997} for the first reaction in equation \eqref{eqn:h2rx} to be our ionization cross-section for \htwo\ ionization:
\begin{equation}
\sigma^{+N}_{\text{\htwo}}(\nu)=\begin{cases}
0,&h\nu<15.42\\
6.2\times10^{-18}h\nu-9.4\times10^{-17},&15.42\le h\nu<16.50\\
1.4\times10^{-18}h\nu-1.48\times10^{-17},&16.50\le h\nu<17.7\\
2.5\times10^{-14}(h\nu)^{-2.71},&17.7\le h\nu,
\end{cases}
\label{eqn:h2ion_cs}
\end{equation}
where the units for $\sigma^{+N}_{\text{\htwo}}(\nu)$ are in cm$^2$ and $h\nu$ are in eV. The continuous function, $\sigma^{+N}_{ \text{\htwo}}(\nu)$, in equation \eqref{eqn:h2ion_cs} is replaced in equation \eqref{eqn:h2ionrate} by discrete values, $\sigma^{N}_{i\text{\htwo}}$, that are the average cross-sections over each photon group, $i$.

We treat ionization by cosmic rays as a constant rate, using for the primary interaction $\xi_{\text{\htwo}}^{\text{prime}}=3.5 \times 10^{-16} \text{ s}^{-1}$ as measured by \citet{Indriolo2012} and $\xi_{\text{\hi}}^{\text{prime}}=1.78 \times 10^{-16} \text{ s}^{-1}$ as measured by \citet{Indriolo2015}. When cosmic rays ionize \htwo\, the resulting H$^+_2$ molecule either becomes two hydrogen atoms by dissociative recombination or transfers its charge to a hydrogen atom, leading to \htwo+\hii\ \citep{Indriolo2012}. However, just as above where we treat \htwo\ photoionization as a dissociation, we will also treat cosmic ray ionization of \htwo\ as a straight dissociation. The extra factors of 2.15 and 2.5 in equations \eqref{eqn:crh2} and \eqref{eqn:crhi} account for secondary ionization, where fast-moving electrons from the first cosmic ray ionization rapidly ionize more gas. We use the methodology developed by \citet{Gong2017} and inspired by \citet{Glassgold1974} where secondary ionization happens 1.5 times the primary rate for atomic gas and 1.15 for molecular gas. Cosmic ray ionization is not relevant to every environment, and we have left it as optional in \textsc{ramses-rt}.

\subsubsection{Molecular hydrogen self-shielding by line overlap}
\label{sec:ss}
\textsc{Ramses-RT} already includes shielding for all species by destruction of the photons that ionize or dissociate the gas and dust shielding. However, as described in our introduction, we need to enhance the destruction of LW photons due to \htwo\ self-shielding processes. Here more photons are absorbed than \htwo\ destroyed, and LW line overlap interferes with dissociation at higher column densities. 

We take advantage of \textsc{Ramses-RT}'s unique position as a moment-based radiative transfer code. LW absorption is highly dependent on the photon's wavelength. We do not track individual bands of the LW photons, but as a group we can determine an overall reduction in photon number density because of \htwo\ absorption. We introduce self-shielding in the destruction term for the photon density update to determine how many photons are absorbed by \htwo,
\begin{equation}
D_{1\text{\htwo}}=S^s_{1\text{\htwo}} c_{\text{r}}\sigma^N_{1\text{\htwo}}n_{\text{\htwo}},
\label{eqn:ss}
\end{equation}
where $D_{1\text{\htwo}}$ is the destruction rate of LW photons, $S^s_{1\text{\htwo}}$ is the self-shielding factor, and $\sigma^N_{1\text{\htwo}}$ is the photodissociation cross-section between \htwo\ and the LW photon band. As we will show in Section \ref{sec:1d}, a constant $S^s_{1\text{\htwo}}\sim$400 reproduces realistic self-shielding in a variety of environments, while being computationally expedient. Our method to enhance LW photodestruction is in contrast to other codes \citep{Gnedin2011,Lupi2018,Katz2017} that calculate \htwo\ self-shielding by decreasing \htwo\ destruction.

Because we keep track of the photon density using a moment-based method, the LW band's cumulative reduction as it travels though an \htwo\ region over multiple time-steps naturally reflects its encounters with a shielding column density; we need neither to convert our volume into a column density, nor to use a non-local method to calculate column density.

Linguistically, our usage of `self-shielding factor' differs slightly from the traditional sense of the term. Other works use a self-shielding factor in front of the \htwo\ photodissociation term, and hence this factor is between 1 and 0, decreasing with column depth. Our self-shielding factor is instead in front of the $N_1$ photodissociation term, and is in this sense an inverse of the traditionally defined factor, being greater than 1.

Multiplying equation \eqref{eqn:h2diss_cs} by our self-shielding factor corresponds to an effective cross-section of $\sim 8\times 10^{-17}$ cm$^2$, or an equivalent column density of $10^{16}$ cm$^{-2}$. In full galaxy simulations, typical cell column densities in molecular regions reach about $10^{19}$ to $10^{23}$ cm$^{-2}$. The smallest column densities in our one-dimensional tests in Section \ref{sec:tests} will range from $10^{16}$ to $10^{18}$ cm$^{-2}$.

\subsection{Thermochemistry step}
\label{sec:thermostep}

Much of the mechanics of the thermochemistry step are detailed in \citet{Rosdahl2013}, with additions from \citet{Rosdahl2015a}. Here we emphasize our \htwo\ addition to the formalism.

Over a time-step, from $t$ to $t+\Delta t$, \textsc{Ramses-RT} evolves the thermochemistry state in each cell given by  $ \mathcal{U_T}=(\varepsilon,x_{\text{\hi}}, \allowbreak x_{\text{\hii}},x_{\text{\hei}}, x_{\text{\heii}}, N_i, \textbf{F}_i)$, where $\varepsilon=E-\frac{1}{2}\rho\textbf{u}^2$ is the thermal energy density. The non-equilibrium thermochemistry equations are too stiff to be solved expediently by an implicit solver, and instead are solved in a fixed order as inspired by \citet{Anninos1997}. The order in which the equations are solved is as follows: photon density and flux update, thermal update, hydrogen fraction update, and helium ionization fraction update. At the end of each step, the quantity is checked to see if it has changed more than 10 per cent. If it has then there is no update and the procedure is run again with $0.5 \Delta t$. Once every quantity has been updated and the 10 per cent change has not been violated, a final check is taken. If the change in $ \mathcal{U_T}$ is less than 5 per cent then the next time-step will be $2\Delta t$. 

The following subsections detail each quantity in the thermochemistry step. The equations are given for case A recombination, but if OTSA is used then case B recombination rates will replace them.

\subsubsection{Photon density and flux update}
\label{sec:phoup}
The photon density, $N$, and flux, $F$, are updated by each photon group, $i$, individually since they operate independently of each other. They are given by
\begin{align}
\frac{\partial N_i}{\partial t}&=\dot{N}_i+C_i-N_iD_i,\\
\frac{\partial F_i}{\partial t}&=\dot{F}_i-F_iD_i,
\label{eqn:dnf}
\end{align}
where $\dot{N}_i$ is the change in photon density from the RT transport solver, $C_i$ is the photon creation from recombination, $D_i$ is photon destruction from absorption terms, and $\dot{F}_i$ is the change in photon flux. There is no corresponding creation term for the flux because radiation from recombination is assumed to be spherically symmetric. 

Creation and destruction are given by
\begin{align}
C_i&=\sum^{\text{\hii, \hei, \heii}}_{j}b^{rec}_{ji}(\alpha^A_j(T)-\alpha^B_j(T))n_j n_e, \label{eqn:phocr}\\
D_i&=\kappa \rho Z f_{\text{d}} c_{\text{r}} + A^{PE}_i(T)+\sum^{\text{\htwo, \hi, \hei, \heii}}_{j} S^s_{ij} c_{\text{r}} \sigma^N_{ij}n_j, 
\label{eqn:phode}
\end{align}
where $b^{rec}_{ji}$ is a boolean to describe the photon group that the $j$-species recombines into, $\alpha^A_j(T)$ and $\alpha^B_j(T)$ are the case A and B recombination rates, $n_j$ is the number density of gas $j$, $n_e$ is the number density of electrons, $\kappa$ is the dust opacity ($\kappa_{\text{P}}$ for $N_i$ and $\kappa_{\text{R}}$ for $F_i$), $\rho$ is the gas volume density, $f_{\text{d}}$ is the fraction of gas that holds dust, $c_{\text{r}}$ is the reduced speed of light, and $\sigma^N_{ij}$ is the destruction cross-section between gas species $j$ and photon group $i$. $S^s_{ij}$ is the self-shielding factor for \htwo\ as described in Section \ref{sec:molrecipe} to boost the destruction of LW photons. If the photon species is LW and the gas species is \htwo\ then $S^s_{1\text{\htwo}}=400$; otherwise $S^s_{ij}=1$. 

$A^{PE}_i$ is the the absorption term from the photoelectric effect \citep{Bakes1994,Wolfire2003}, which we expand on in Section \ref{sec:thermup}. It is only non-zero in the LW band:
\begin{equation} 
A^{PE}_1(T)=8.125\times10^{-22} \text{cm}^2 \epsilon_{\text{ff}}(T) c_{\text{r}} n_H Z f_{\text{d}},
\label{eqn:ape}
\end{equation}
where for the cross-section we divide the heating rate from equation \eqref{eqn:peh} by the Habing field \citep{Habing1968} and $\epsilon_{\text{ff}}$ is given by equation \eqref{eqn:peheff}. The photoelectric effect occurs over energies 8 to 13.6 eV, which goes a little lower than the LW band, but we do not add an extra photon group for computational expediency.

If OTSA is on then there is no creation term (equation \ref{eqn:phocr}) because the photons are assumed to be immediately reabsorbed.  \htwo\ formation by dust does not involve the emission of photons, and while the gas phase formation does, its rate is much too weak to have an impact on our simulations, and therefore the $C_i$ term does not involve any photons from \htwo\ creation. 

Photon density and flux advance in time with a partly semi-implicit Euler formulation given by,
\begin{align}
    N_i^{t+\Delta t} &=\frac{N_i^t+\Delta t(\dot{N}_i+C_i)}  {1+\Delta t D_i}, \\
    F_i^{t+\Delta t} &=\frac{F_i^{t}+\Delta t \dot{F}_i} {1+\Delta t  D_i}.
\end{align}

At the end of this step momentum is transferred from the photons to the gas and the energy absorbed by dust is added to the IR photons if this group is in use, as outlined in \citet{Rosdahl2015a}.

\subsubsection{Thermal update}
\label{sec:thermup}
For each gas cell in \textsc{Ramses-RT}, the temperature can be obtained via
\begin{equation}
T=\varepsilon\frac{(\gamma-1)m_H}{\rho k_{\text{B}}}\mu,
\label{eqn:ramtemp}
\end{equation}
where $\varepsilon$ is the thermal energy density, $\gamma$ is the ratio of specific heats, $m_H$ is the proton mass, $\rho$ is the density, $k_{\text{B}}$ is the Boltzman constant, and $\mu$ is the average gas particle mass in units of $m_H$.

However, because $\mu$ depends on the ionization fraction, $T_{\mu}=T/\mu$ is evolved instead of $T$ via,
\begin{align}
\label{eqn:dtempdt}
 \frac{\partial T_{\mu}}{\partial t}&=\frac{(\gamma-1)m_H}{\rho k_B}(\mathcal{H}-\mathcal{L}),\\
 \mathcal{H}&=\sum^{\text{\htwo, \hi, \hei, \heii}}_{j}n_j\sum^M_{i=1}c_{\text{r}}N_i(\bar{\epsilon}_i\sigma^E_{ij}-\epsilon_j\sigma^N_{ij})\\
&+\mathcal{H}_{\text{PE}}(T)+\mathcal{H}_{\text{UVP}}(T)+\mathcal{H}_{\text{\htwo}}(T)+\mathcal{H}_{\text{CR}}(T),\nonumber\\  
\mathcal{L}&= \left[ \zeta_{\text{\hi}}(T) + \psi_{\text{\hi}}(T) \right] \ n_e \ n_{\text{\hi}} \\
   &+ \zeta_{\text{\hei}}(T) \ n_e \ n_{\text{\hei}}  \nonumber \\
   &+ [ \zeta_{\text{{\heii}}}(T) + \psi_{\text{\heii}}(T) + \eta^{\rm{A}}_{\text{\heii}}(T) \nonumber \\
   &+ \omega_{\text{\heii}}(T) ] n_e n_{\text{\heii}}  \nonumber \\
   &+ \eta^{\rm{A}}_{\text{\hii}}(T) \ n_e \ n_{\text{\hii}}  \nonumber \\
   &+ \eta^{\rm{A}}_{\text{\heiii}}(T) \ n_e \ n_{\text{\heiii}}  \nonumber \\
   &+ \theta(T) \ n_e ( n_{\text{\hii}}+n_{\text{\heii}}+4n_{\text{\heiii}} )  \nonumber \\
   &+ \varpi(T) \ n_e \nonumber \\
   &+\Lambda_Z(T) \nonumber \\
   &+\Lambda_{\text{\htwo}}(T). \nonumber
\end{align} 
In the heating term, $\mathcal{H}$, $\bar{\epsilon}_i$ is the photon average energy, $\epsilon_j$ is the photodestruction energy, $\sigma^N_{ij}$ is the average cross-section between group $i$ and species $j$, and $\sigma^E_{ij}$ is the energy-weighted cross-section. In simulations with star particles, these are calculated from SED tables. For this paper, we do not work with stars and instead use the cross-sections averaged over a black body. $\mathcal{H}_{\text{PE}}(T)$ is heating from the photoelectric effect, $\mathcal{H}_{\text{UVP}}(T)$ is heating from UV pumping, $\mathcal{H}_{\text{\htwo}}(T)$ is heating from \htwo\ formation, and $\mathcal{H}_{\text{CR}}(T)$ is heating from cosmic ray ionization. The cooling term, $\mathcal{L}$, includes  collisional ionization $\zeta$, collisional excitation $\psi$, recombination $\eta$, dielectronic recombination $\omega$, Bremsstrahlung $\theta$, and Compton cooling $\varpi$. Their functional forms and sources are given in \citet{Rosdahl2013}. $\Lambda_Z$ is the contribution of metals to cooling, from tables generated by \textsc{cloudy} above $10^4$ K, and below $10^4$ K using the fine-structure cooling rates from \citet{Rosen1995} \citep{Rosdahl2017}. \textbf{$\Lambda_{\text{\htwo}}$ is cooling from \htwo.}

We discuss in the following paragraphs the heating and cooling processes added for the \htwo\ chemistry, while the remainder are given in  \citet{Rosdahl2013}.

$\mathcal{H}_{PE}$ is the heating from the photoelectric effect, as given by \citet{Bakes1994} and updated by \citet{Wolfire2003}:
\begin{align}
\mathcal{H}_{\text{PE}}(T)&=1.3\times 10^{-24} \text{ erg s}^{-1} \epsilon_{\text{ff}}(T) G_0 n_H Z f_{\text{d}} \label{eqn:peh}\\
G_0&=\epsilon_1 N_1 c_{\text{r}}/(1.6\times 10^{-3}\text{erg s}^{-1}\text{cm}^{-2}), \label{eqn:g0}\\
\epsilon_{\text{ff}}(T)&=\frac{4.87 \times 10^{-2}}{1+4\times 10^{-3} (G_0 \sqrt{T_{\text{K}}}/(0.5n_e))^{0.73}} \label{eqn:peheff}\\
&+\frac{3.65 \times 10^{-2} (T_{\text{K}}/10^4)^{0.7}}{1+2\times 10^{-4} (G_0 \sqrt{T_{\text{K}}}/(0.5n_e))}, \nonumber
\end{align}
where $\epsilon_1$ is the energy of group 1 in erg, $G_0$ normalizes our field to the Habing value \citep{Habing1968}, $\epsilon_{\text{ff}}$ is the photoelectric heating efficiency, and $T_K$ is the temperature in Kelvin.

Heating from UV pumping is a result of LW photon absorption by \htwo\ that does not lead to dissociation of the molecule, but nonetheless heats it. We use the prescription by \citet{Baczynski2015} based on calculations by \citet{Draine1996} and \citet{Burton1990}:
\begin{align}\label{eqn:huvp}
\mathcal{H}_{\text{UVP}}(T)&=2.22\times 10^{-11} \text{ erg } N_{1} \sigma^N_{11}c_{\text{r}} n_{\text{\htwo}} \\
&\times \frac{C_{\text{dex}}(T)}{C_{\text{dex}}(T)+2\times 10^7\text{ s}^{-1}} , \nonumber\\
C_{\text{dex}}(T)&=10^{-12}\Big(1.4 \text{e}^{-18100/(T_{\text{K}}+1200)}x_{\text{\htwo}}  \\ &+\text{e}^{-1000/T_{\text{K}}}x_{\text{\hi}}\Big)\sqrt{T_{\text{K}}}n_{\text{H}} \text{ s}^{-1}, \nonumber
\end{align}
where $C_{\text{dex}}$ represents the collisional deexcitation rate.

When \htwo\ forms, it releases a small amount of heat depending on the formation mechanism. For formation on dust grains and gas phase formation, we use the formulation by \citet{Hollenbach1979} and for formation by three-body collisions we use \citeauthor{Omukai2000}'s (\citeyear{Omukai2000}) formulation:
\begin{align}\label{eqn:hh2}
\mathcal{H}_{\text{\htwo}}(T)&= 1.6022\times 10^{-12} \text{ erg}\\ 
&\times\Big( (0.2 + 4.2/(1+n_{cr}(T)/n_H)) \alpha_{\text{\htwo}}^{\text{Z}}Z f_{\text{d}} n_H n_{\text{\hi}}\nonumber\\
&+ 3.53/(1+n_{cr}(T)/n_H)\alpha_{\text{\htwo}}^{\text{GP}}n_{\text{\hi}}n_{\text{e}} \nonumber\\
&+ 4.48/(1+n_{cr}(T)/n_H)\beta_{\text{3B}}n_{\text{\hi}}^2(n_{\text{\hi}} + n_{\text{\htwo}}/8) \Big) \nonumber \\
n_{cr}(T) &= 10^6 \times T_{\text{K}}^{-0.5}/\Big(1.6x_{\text{\hi}}\text{e}^{-400/{T_{\text{K}}}^2} \\
&+1.4x_{\text{\htwo}}\text{e}^{-12000/(T_{\text{K}}+1200)}\Big) \text{ cm}^{-3} \nonumber,
\end{align}
where $n_{cr}$ is the critical density.

We find that in our regimes of interest, UV pumping and \htwo\ formation contribute negligible heating compared to the other processes, but we nonetheless include them for completeness.

Our final heating term is from cosmic ray ionization. While this heating rate does depend on gas density, we opt for a simple approximation and use the measurements of \citet{Glassgold2012} in which the average event deposits about 10 eV of energy:
\begin{align}\label{eqn:hcr}
\mathcal{H}_{\text{CR}}(T)&=1.6022 \times 10^{-11} \text{ erg } (\xi_{\text{\hi}}n_{\text{\hi}} \\
&+\xi_{\text{\htwo}}n_{\text{\htwo}}+ 1.1\xi_{\text{\hi}}n_{\text{\hei}}), \nonumber
\end{align}
where for the cosmic ray ionization rate of helium we use $\zeta_{\text{\hei}}=1.1\zeta_{\text{\hi}}$ \citep{Glover2010}.

Below temperatures of 5000 K, \htwo\ is the dominant coolant \citep{Gnedin2011}. We use a cooling function that is similar in form to \citet{Halle2013} for \hi-\htwo\ and \htwo-\htwo\ collisional cooling only in the low-density limit ($n\rightarrow0$), because our galactic simulations will not resolve high-enough densities to reach local thermal equilibrium rates (LTE),
\begin{equation}
\Lambda_{\text{\htwo}}(T)=\Lambda_{\text{\htwo\hi}(n\rightarrow0)}(T)n_{\text{\hi}}n_{\text{\htwo}}+\Lambda_{\text{\htwo\htwo}(n\rightarrow0)}(T)n_{\text{\htwo}}^2,
\label{eqn:lambda}
\end{equation}
where $\Lambda_{\text{\htwo\hi}(n\rightarrow0)}$ and $\Lambda_{\text{\htwo\htwo}(n\rightarrow0)}$ are the low-density limits of the \htwo\ collisional cooling coefficients from \citet{Hollenbach1979} in units of cm$^3$erg s$^{-1}$.

The temperature is then updated semi-implicitly using the updated values for photon density and flux from Section \ref{sec:phoup}, but the un-updated values for the hydrogen and helium species:
\begin{equation}
    T_{\mu}^{t+ \Delta t}= T_{\mu}^{t} + \frac{\Lambda K \Delta t}{1-\Lambda' K \Delta t}.
    \label{eqn:tupdate}
 \end{equation}
Here $\Lambda \equiv \mathcal{H}-\mathcal{L}$, $\Lambda' \equiv -\frac{\partial \mathcal{L}}{\partial T_{\mu}}$, and $K \equiv \frac{(\gamma-1)m_H}{\rho k_B}$.

\subsubsection{Species fraction update}
We only store the variables $x_{\text{\hi}}$ and $x_{\text{\hii}}$, and recover $x_{\text{\htwo}}$ via $x_{\text{\htwo}}=0.5(1-x_{\text{\hi}}-x_{\text{\hii}})$. However, all three quantities are evolved in order to ensure consistency and stability. 

These fractions evolve as,
\begin{align}
\frac{\partial {x}_{\text{\htwo}}}{\partial t}&=x_{\text{\hi}}\Big( \alpha_{\text{\htwo}}^{\text{Z}} Z f_{\text{d}} n_{\text{H}} + \alpha_{\text{\htwo}}^{\text{GP}} n_{\text{e}} \label{eqn:h2update}\\
&+ \beta_{\text{3B}}n_{\text{\hi}}(n_{\text{\hi}}+n_{\text{\htwo}}/8) \Big) \nonumber \\
&-x_{\text{\htwo}} \Big(\beta_{\text{\htwo\hi}}n_{\text{\hi}}+\beta_{\text{\htwo\htwo}}n_{\text{\htwo}} \nonumber\\
&+\xi_{\text{\htwo}}+\sum^M_{i=1}\sigma^N_{i\text{\htwo}}c_{\text{r}}N_i\Big), \nonumber \\
\frac{\partial {x}_{\text{\hi}}}{\partial t}&=2x_{\text{\htwo}} \Big(\beta_{\text{\htwo\hi}}n_{\text{\hi}}+\beta_{\text{\htwo\htwo}}n_{\text{\htwo}} \label{eqn:hiupdate} \\
&+\xi_{\text{\htwo}}+\sum^M_{i=1}\sigma^N_{i\text{\htwo}}c_{\text{r}}N_i \Big) \nonumber \\
&+ x_{\text{\hii}}\alpha^A_{\text{\hii}} n_{\text{e}} \nonumber \\
&-x_{\text{\hi}}\Big( 2\alpha_{\text{\htwo}}^{\text{Z}} Z f_{\text{d}} n_{\text{H}} + 2\alpha_{\text{\htwo}}^{\text{GP}} n_{\text{e}} \nonumber \\
&+2\beta_{\text{3B}}n_{\text{\hi}}(n_{\text{\hi}}+n_{\text{\htwo}}/8) \nonumber \\ &+\beta_{\text{\hi}}n_{\text{e}}+\xi_{\text{\hi}}+\sum^M_{i=1}\sigma^N_{i\text{\hi}}c_{\text{r}}N_i \Big), \nonumber \\
\frac{\partial {x}_{\text{\hii}}}{\partial t}&=x_{\text{\hi}} \Big(\beta_{\text{\hi}}n_{\text{e}}+\xi_{\text{\hi}}+\sum^M_{i=1}\sigma^N_{i\text{\hi}}c_{\text{r}}N_i \Big) \label{eqn:hiiupdate} \\
&-x_{\text{\hii}}\alpha^A_{\text{\hii}} n_{\text{e}}. \nonumber
\end{align}  

The respective destruction and creation coefficients corresponding to \htwo\ are given in Section \ref{sec:ratecoeff}, while the coefficients for \hi\ and \hii\ are given in \citet{Rosdahl2013}. Each of these equations \eqref{eqn:h2update} to \eqref{eqn:hiiupdate} for a species fraction $x$ may be reformulated as, 
\begin{equation}
\frac{\partial x}{\partial t}=C-xD,
\label{eqn:xdiff}
\end{equation}
for their creation term $C$ and destruction $D$. We then update each species fraction in order of \htwo, \hi, and \hii\ using the semi-implicit method,
\begin{equation}
x^{t+\Delta t}=\frac{x^t+C\Delta t}{1+D\Delta t}.
\label{eqn:xpi}
\end{equation}
This expression always uses the updated values of $N_i$ and $T_{\mu}$ from Sections \ref{sec:phoup} and \ref{sec:thermup}. The \htwo\ update uses entirely un-updated values of the species fractions. The \hi\ update uses the new value for \htwo, while all other species fractions are un-updated. Finally, the \hii\ update uses the new \hi\ fraction. 

At the end of this step, we enforce conservation of hydrogen, by checking that $2x_{\text{\htwo}}+x_{\text{\hi}}+x_{\text{\hii}}=1$, and when this fails we lower the largest fraction accordingly. 

Updating the fraction of helium species, between \hei, \heii, and \heiii, follows an almost identical procedure as above and is unchanged from \citet{Rosdahl2013} with one exception. We provide the option of using cosmic ray ionization of \hei.

\section{The Tests}
\label{sec:tests}

\citet{Rosdahl2013} use the Iliev series of benchmark tests \citep{Iliev2006, Iliev2009} for radiative transfer codes in atomic and ionized environments to verify its robustness. It is difficult to create tests with analytical solutions for radiative transfer codes, and so instead radiative transfer codes are compared to each other in these benchmark tests. If many codes agree, then they are taken to be correct.

For \htwo\ formation in galaxy codes, however, there is no series of benchmark tests. Instead, we compare our code to PDR codes optimized for smaller scales. Our strategy is to begin with simple zero-dimensional tests, and add increasing complexity. In zero dimensions, that is a single cell, we can compare to an analytic solution. For one dimension, we can compare to more detailed PDR codes that are specialized for these scales, and extrapolate the results to two and three dimensions. Finally, for three dimensions we introduce a Str{\"o}mgren sphere for an ionized hydrogen front in a neutral medium shell protecting a larger molecular medium. Our aim is to test our methodology against known solutions where they exist, and ensure sensible outcomes where there are no known solutions. 

All these tests use only hydrogen without helium, and frozen hydrodynamics. For the LW band we use groups 1, 11.2 $\le h\nu \le$ 13.6 eV, and for \htwo\ and \hi\ ionization radiation we use group 2, 13.6 $\le h\nu \le$ 24.59 eV. Higher energies also ionize hydrogen, but their cross-sections are small enough to not have an impact.

\subsection{Single cell convergence}
\label{sec:singlecell}
These zero-dimensional tests are similar to those run in \citet{Rosdahl2013} to see if our method for \htwo\ thermochemistry makes sense in simple situations. For all scenarios it is important to test for smoothness of evolution and if the final state is physically sensible. We evolve single cells with a homogeneous radiation-gas fluid. They have a range of hydrogen densities, initial temperatures, and initial atomic/ionized fractions. Density is fixed while hydrogen fractions evolve over time. Metallicity is fixed at the Solar value. We run each cell for $2\times 10^2$ Myr, which is a little longer than the possible lifetime for molecular clouds from 30 to 100 Myr \citep{Zasov2014}.

There are four scenarios: with and without a fixed UV radiation field and with a fixed temperature or variable temperature. In the fixed-temperature cases, we need to see if the cell's hydrogen fractions evolve to the equilibrium value. We obtain the equilibrium value by numerically iterating over the rate equations \eqref{eqn:hiirate}, \eqref{eqn:hirate}, and \eqref{eqn:h2rate} until a steady state is reached. 

For each scenario we test a grid of six fixed hydrogen densities ($10^{-4} \le  n_H \le 10^6 \eqcnc$) and five fixed/initial temperatures ($10 \le T \le 10^7 $ K). In addition, for each density and temperature combination we test initial fractions of $x_{\text{\hi}}=1.0, 0.8, 0.5, 0.2,$ and 0.0. The initial molecular fraction is always zero. The UV field is calculated from the $z=0$ Haardt and Madau background \citep{Haardt1996} over the \htwo-dissociating and \hi-ionizing photon groups at redshift zero (photoionization: $\Gamma_{\text{UV\htwo}}=2.6\times10^{-18}$s$^{-1}$ and $\Gamma_{\text{UV\hi}}=3.6\times10^{-14}$s$^{-1}$; and photoheating $\mathcal{H}_{\text{UV\htwo}}=1.8\times10^{-30}$erg s$^{-1}$ and $\mathcal{H}_{\text{UV\hi}}=2.4\times10^{-25}$erg s$^{-1}$). Cosmic ray ionization and heating are off.

\begin{figure*}
\includegraphics[width=\textwidth]{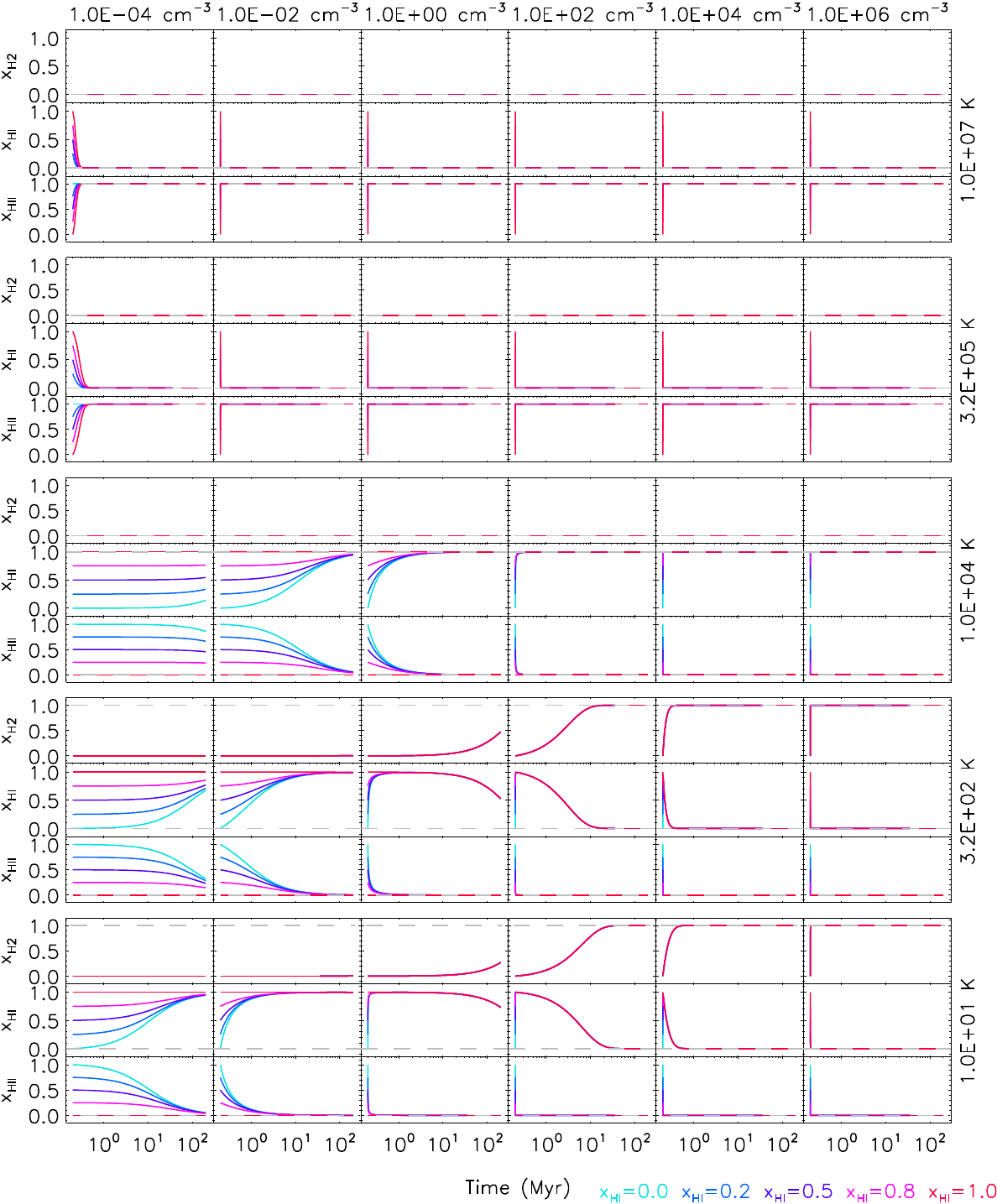}
\caption{Hydrogen species evolution in zero-dimensional convergence tests for a range of fixed densities and temperatures. Coloured lines refer to different initial atomic hydrogen fractions. Grey dashed lines are the equilibrium states.}
\label{fig:xno}
\end{figure*}

We begin with fixed temperature and no UV background radiation (Fig. \ref{fig:xno}) and compare the evolution to the equilibrium state for each density and temperature. In this situation, given any temperature the equilibrium state is the same for all densities. As  expected, higher temperatures of $3.2\times10^5$ K and over lead to an ionized equilibrium state while intermediate $10^4$ K temperatures yield atomic, and temperatures at $3.2\times10^2$ K and lower lead to a molecular state. In lower-density environments below 10$^4$ K, from $10^{-4}$ cm$^{-3}$ to $1$ cm$^{-3}$, the cell does not reach a fully molecular equilibrium state within the simulation time. By contrast, as the density increases the cells reach the equilibrium state much more quickly and \htwo\ can form. In their work on molecular cloud simulations, \citet{Glover2007a} show that in non-turbulent clouds with initial densities of 10 cm$^{-3}$, \htwo\ forms in about 10 Myr. Fig. \ref{fig:xno} supports this.

\begin{figure*}
\includegraphics[width=\textwidth]{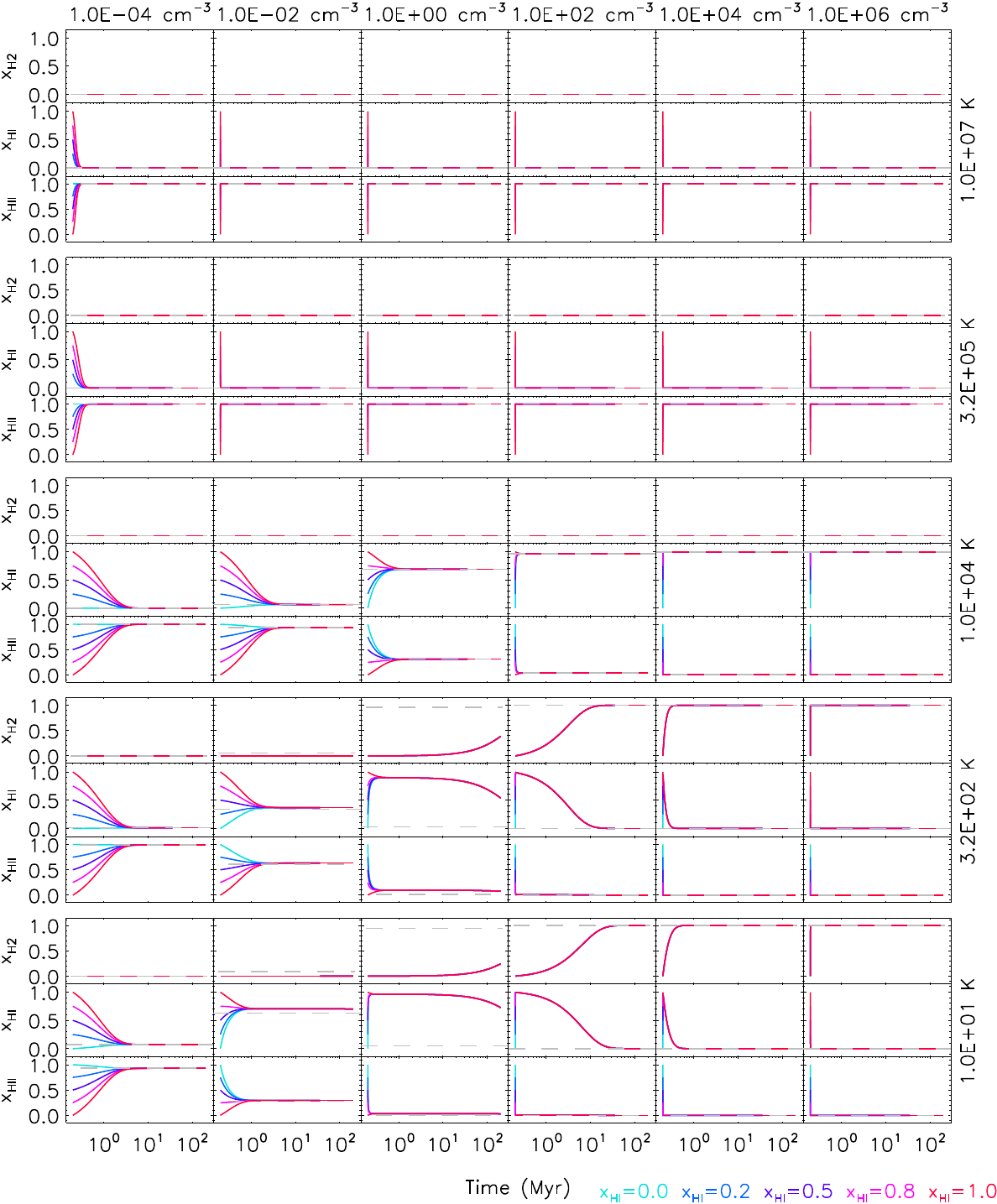}
\caption{Hydrogen species evolution in zero-dimensional convergence tests for a range of fixed densities and temperatures. UV background is on. Coloured lines refer to different initial atomic hydrogen fractions. Grey dashed lines are the equilibrium states.}
\label{fig:xuv}
\end{figure*}

Next, we run the same test again but with a UV  background (Fig. \ref{fig:xuv}). For the same temperatures and densities, the gas is more ionized and less molecular than without a UV background. Now the equilibrium state does depend on both density and temperature. This is because previously we only had the collisional destruction rates that are proportional to the density and so they cancelled out in the equilibrium calculations, while the destruction rate from the UV background is density-independent. $3.2\times10^5$ K and higher yields an entirely ionized state for all densities. At $10^4$ K, the gas is fully ionized at $10^{-4}$cm$^{-3}$. At 100 cm$^{-3}$, the final fraction is mostly \hi, with traces of \hii, and higher densities are entirely atomic. At lower temperatures, $3.2\times10^5$ K and below, our final states are only fully molecular at densities of 100 $\eqcnc$ and higher. At $10^{-2} \eqcnc$ the final state is a \hi\ and \hii\ mix, while at lower densities the cell is almost entirely ionized. Our $10^{-2}$ and 1 $\eqcnc$ cells at these lower temperatures do not reach the equilibrium state in the simulation time.

\begin{figure*}
\includegraphics[width=\textwidth]{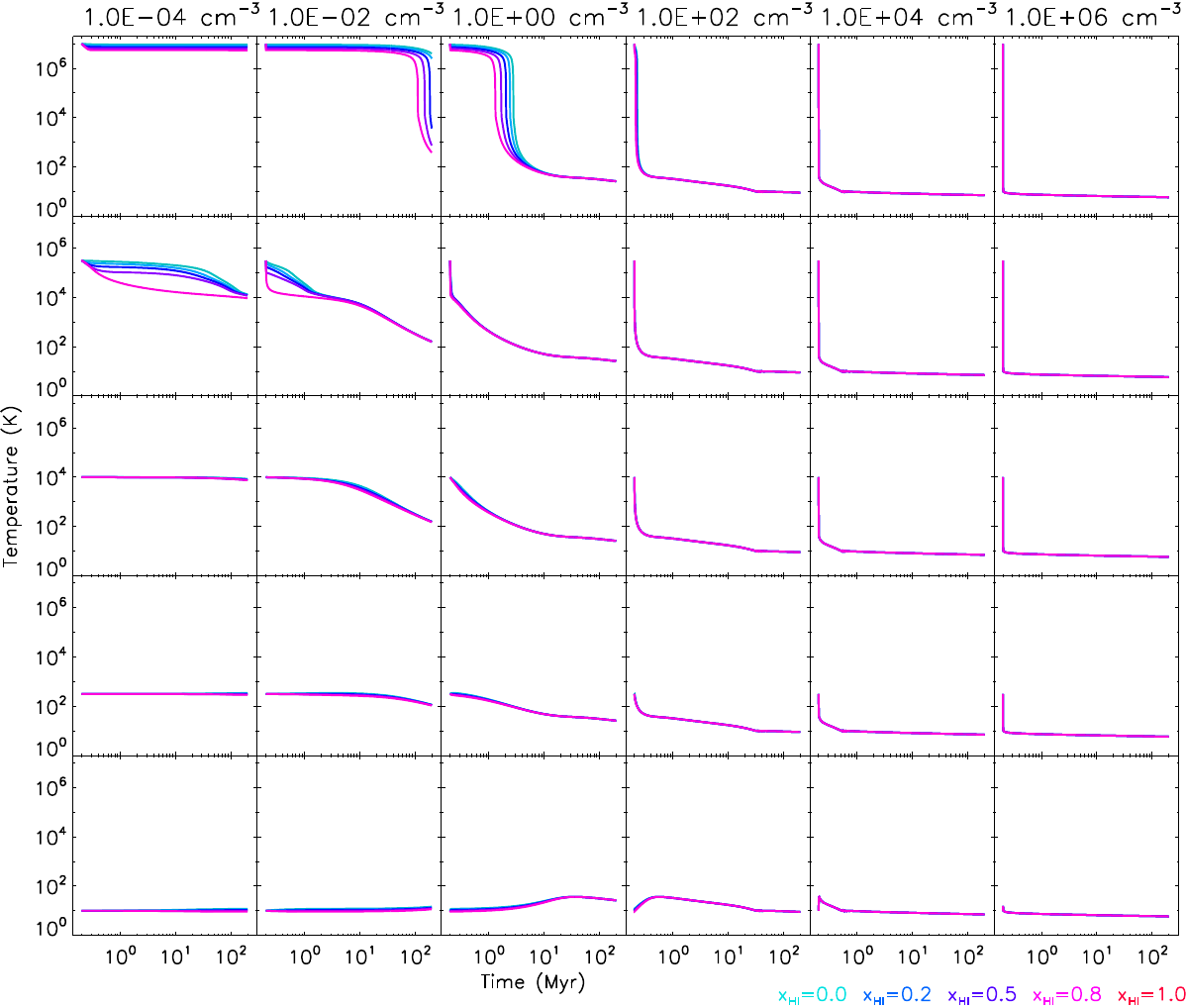}
\caption{Temperature evolution in zero-dimensional tests for a range of fixed densities and initial temperatures. Coloured lines refer to different initial atomic hydrogen fractions. Initial molecular fraction is always zero.}
\label{fig:tno}
\end{figure*}

The next permutation is to allow for a variable temperature, again with the same range of densities and initial temperatures. We rerun this first without a UV background (Fig. \ref{fig:tno}). In these conditions it is clear that little cooling occurs in the lowest-density environment, $10^{-4} \eqcnc$. However as density is increased, the cooling  increases for each initial temperature state. By $10^{2} \eqcnc$ and higher, every initial temperature state cools quickly to a $\sim$10 K floor below which a cell will not cool any further. This number is due to metal cooling (Section \ref{sec:thermup}). The initially 10 K cells change little because they are already at this floor. The necessity of \htwo\ for cooling is clear for the $3.2\times10^5$ K case and higher, where the cooling begins to be affected by the initial \hi\ fraction. The initial entirely atomic states cool the fastest, being the easiest to convert to \htwo, while increasing the initial ionization fraction slows cooling. A certain density of \htwo\ is reached before cooling begins to accelerate.

\begin{figure*}
\includegraphics[width=\textwidth]{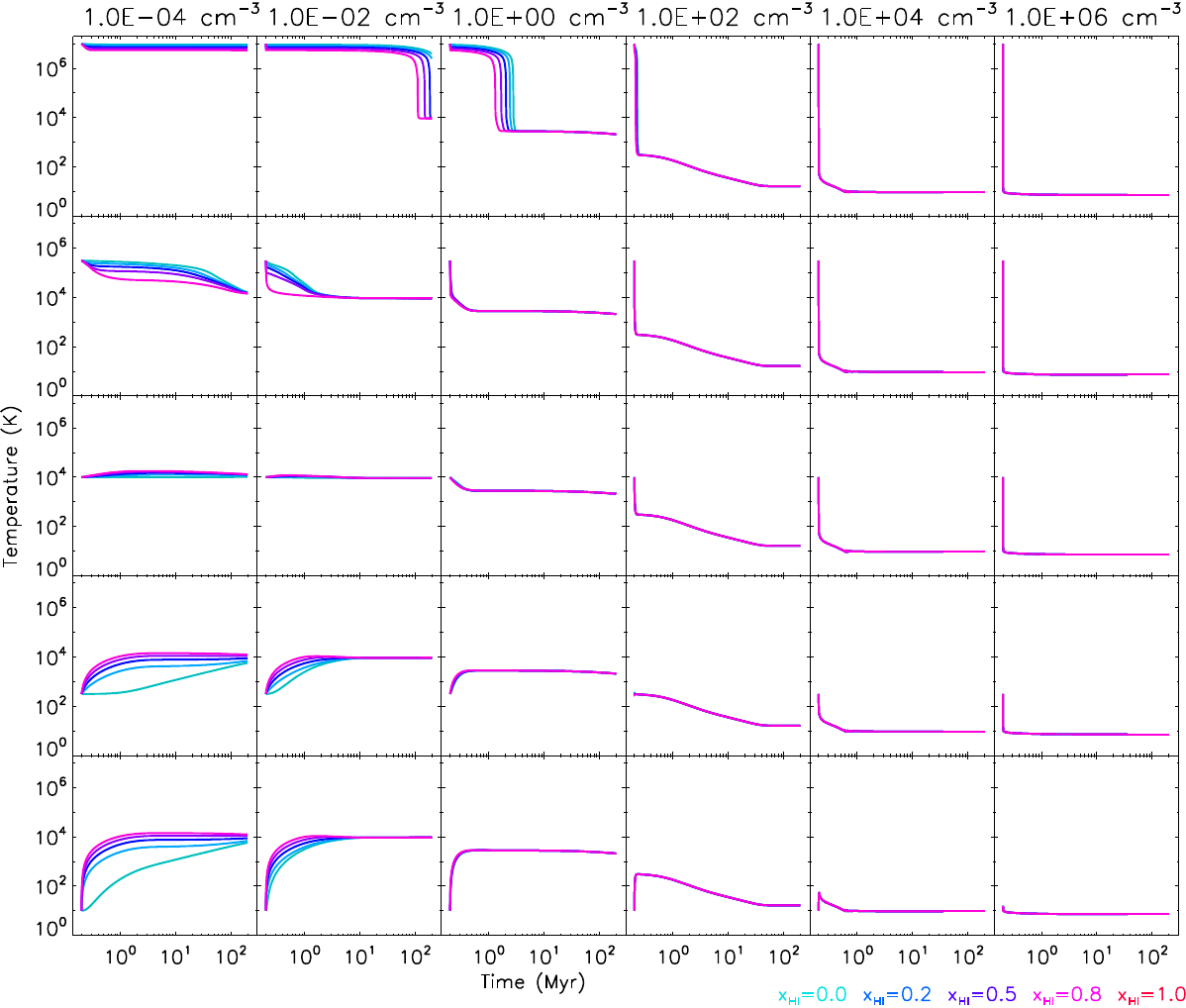}
\caption{Temperature evolution in zero-dimensional tests for a range of fixed densities and initial temperatures. UV background is on. Coloured lines refer to different initial atomic hydrogen fractions. Initial molecular fraction is always zero.}
\label{fig:tuv}
\end{figure*}

The picture is a little different when we use a UV background (Fig. \ref{fig:tuv}). Here the lowest-density cells, $10^{-4}$ and $10^{-2} \eqcnc$, which cannot form any \htwo, either cool down or heat up towards $10^4$ K instead of 10 K as in the case of no UV. The highest-temperature and lowest-density cell remains an exception, being fairly constant. $1 \eqcnc$ cells cool/heat to $\sim$10$^3$ K, until close to the simulation end when cooling begins again due to a little \htwo\ formation.  $100 \eqcnc$ cells cool/heat briefly to on order of $10^2$ K before quickly cooling to the 10 K floor since \htwo\ soon forms at these higher-densities. Cells $10^4 \eqcnc$ and denser cool quickly to $\sim$10 K. We can compare this scenario to the fixed-temperature case with UV (Fig. \ref{fig:xuv}) where molecular hydrogen does not form at all at the lower-densities, and even at 10 K needs $100 \eqcnc$ to form quickly in significant enough quantities within the simulation time. There are some small oscillations in the $10^4 \eqcnc$ column as the temperature hits $\sim$10 K, due to equation stiffness, but these soon dissipate.

These tests are in line with the equilibrium values for the fixed-temperature cases and for the variable-temperature cases the results are reasonable. We can conclude that our thermochemistry is robust in zero dimensions.

\subsection{Self-shielding calibration}

It is important to calibrate our self-shielding factor in equation \eqref{eqn:ss} to realistically reproduce the transition depth between \hi\ and \htwo. We set up a series of one-dimensional simulations where a constant source of LW radiation travels through a low-density region and hits a high-density, \htwo\ region. The high-density region is fixed at $n=1, 10, 100,$ and 1000 $\eqcnc$ for fluxes 0.1, 1, and 10$\chi$, fixed temperature at 50 K (chosen for consistency with \citealt{Roellig2007}), and Solar metallicity. $\chi$ is the Draine flux (1.4 $\times 10^8$ photons cm$^{-2}$s$^{-1}$ in the LW band), the standard UV background for the ISM \citep{Draine1978}. There are no cosmic rays in these tests in order to compare with \citet{Bialy2017} who do not consider them. At densities of $10^3 \eqcnc$, the column density of a single cell is $10^{16}$ cm$^{-2}$. Lower densities have cells with column densities $10^{17}$ to $10^{18}$ cm$^{-2}$.

\begin{figure}
\begin{tabular}{l}
\includegraphics[width=\columnwidth]{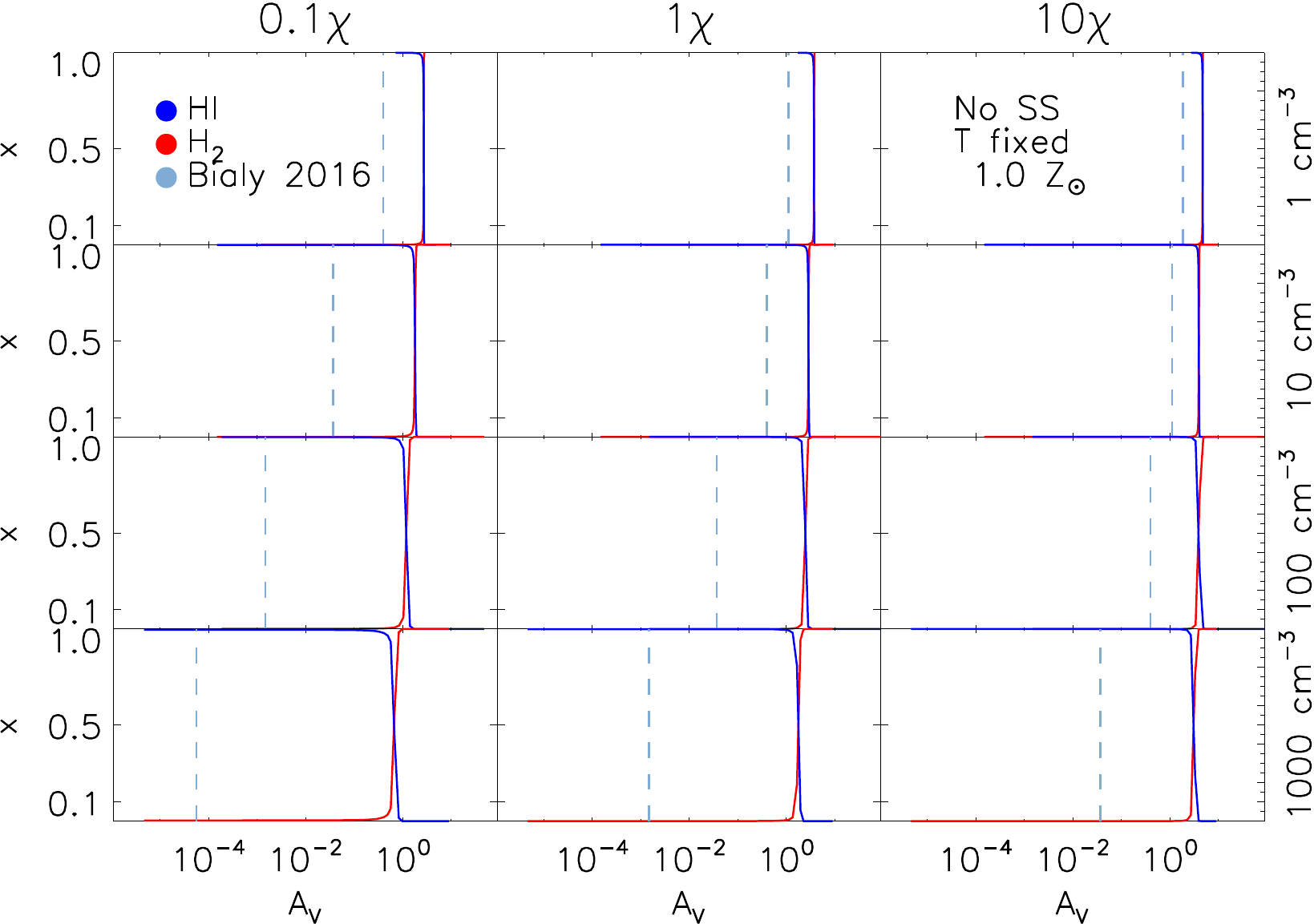}\\
\includegraphics[width=\columnwidth]{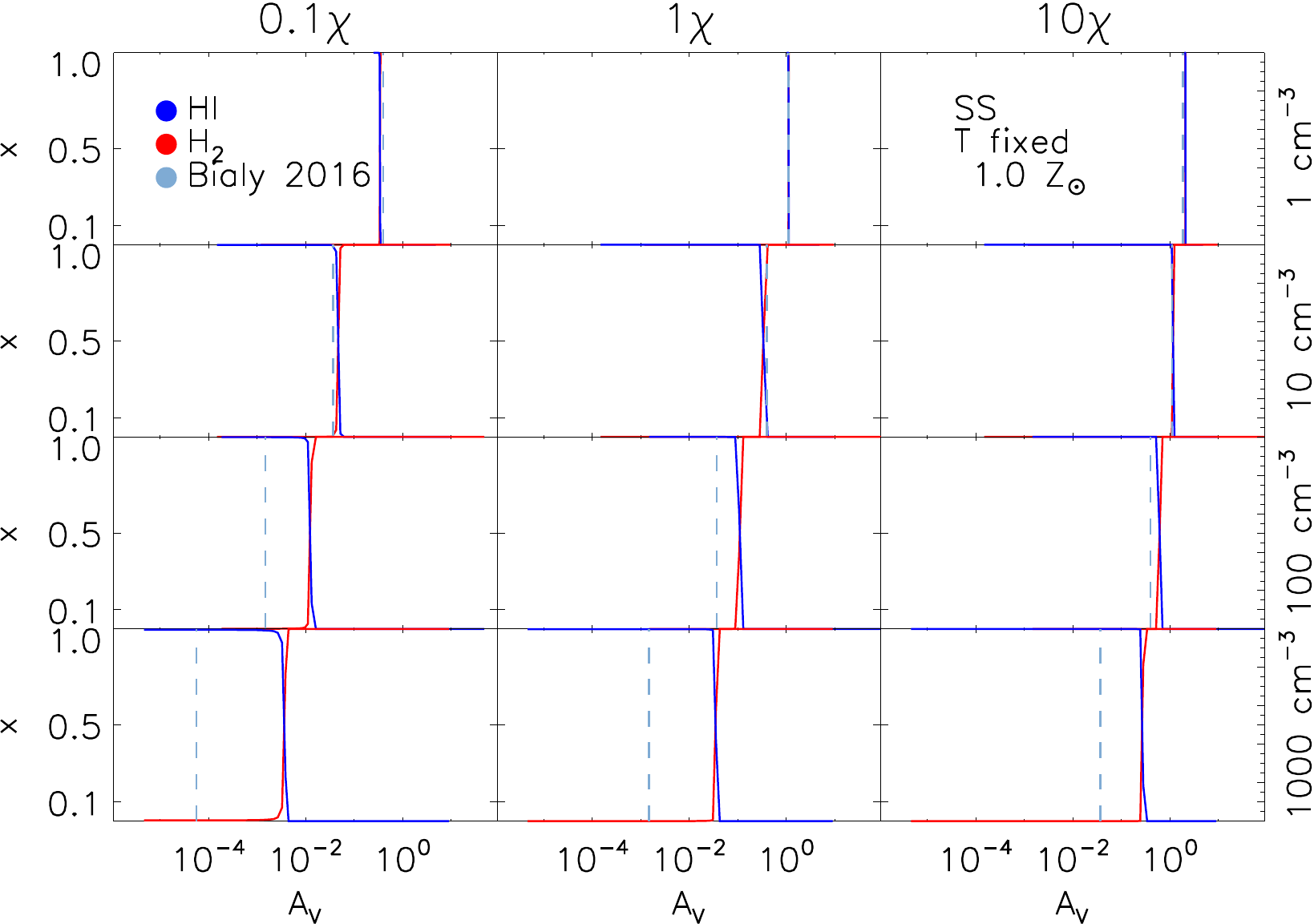}\\
\includegraphics[width=\columnwidth]{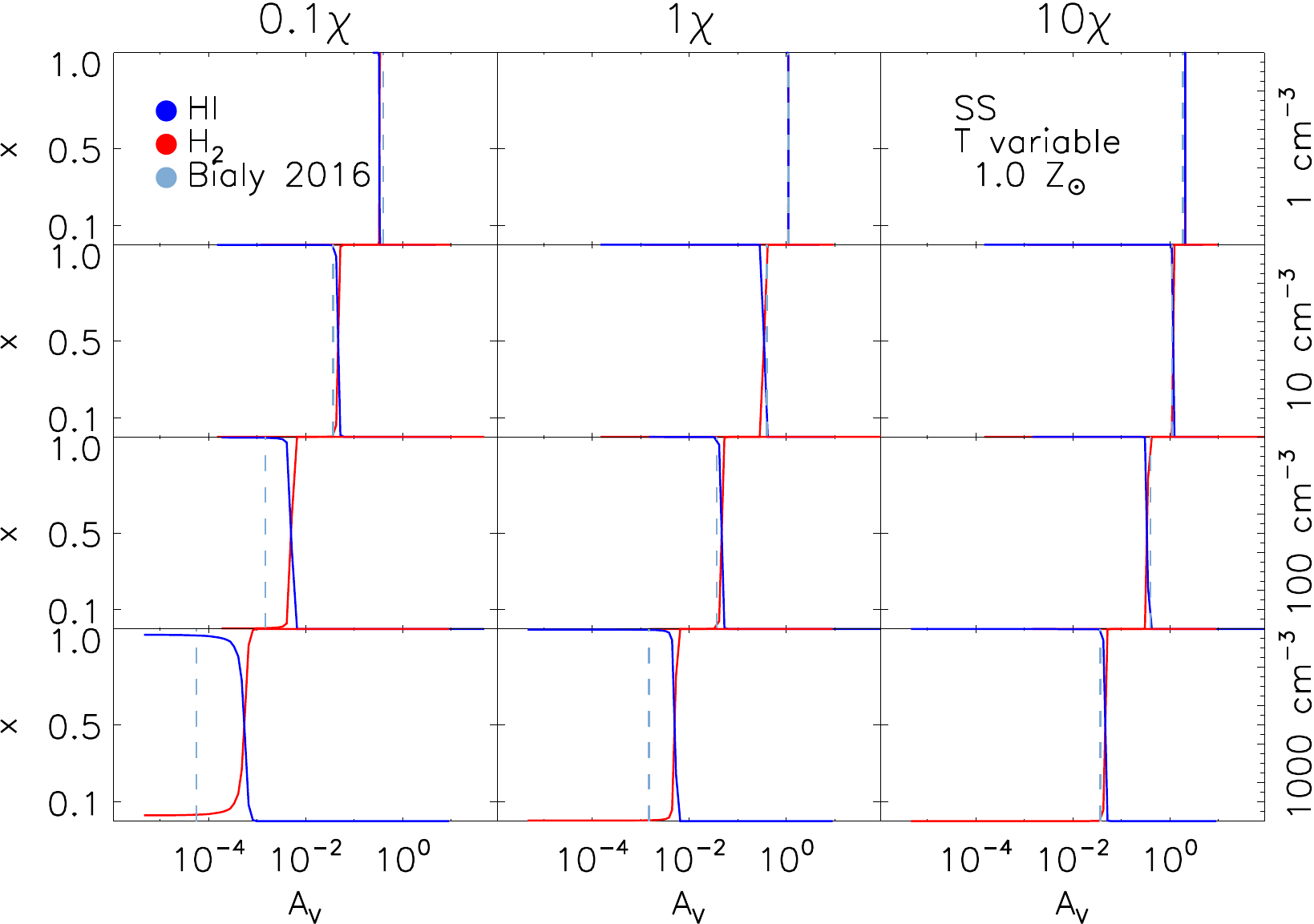}
\end{tabular}
\caption{\hi\ and \htwo\ fraction versus visual extinction in a high-density region hit by LW photons, for a range of fixed densities and fluxes, at Solar metallicity. The theoretical point of transition between \hi\ and \htwo\ is given by the the vertical dotted line calculated from \citeauthor{Bialy2017}'s (\citeyear{Bialy2017}) analytical function. In the top and middle plots, the temperature is fixed at 50 K, while in the bottom plot the temperature is variable. The top plot is without self-shielding, and the bottom two plots use our self-shielding method, given by an enhancement factor of 400 for the LW photodestruction in equation \eqref{eqn:phode}. For the variable-temperature case, we use $T=10$ K to calculate the dotted line, since this is the temperature reached in the molecular region.}
\label{fig:1dss}
\end{figure}

The top plot in Fig. \ref{fig:1dss} shows the transition between \hi\ and \htwo\ in the high-density region, without any self-shielding due to LW absorption line overlap. To convert from column density, $\mathcal{N}_H$, to visual extinction, $A_V$, we use the conversion, $A_V=6.289\times10^{-22}\mathcal{N}_H$ in order to be consistent with \citet{Roellig2007}. We also plot the transition's location as predicted by \citet{Bialy2017}, who give an analytic expression for the column density of transition between \hi\ and \htwo\ based on the \citet{Sternberg2014} theory for PDR regions: 
\begin{align}
\label{eqn:ntrans}
\mathcal{N}_{\text{trans}}&=0.7 \text{ln}\Big[ \Big( \frac{\alpha G}{2} \Big)^{1/0.7}+1\Big]\times \Big(1.9\times10^{-21}Z \text{cm}^2\Big)^{-1}, \\
\alpha G &=0.59F_{LW\chi}\Big( \frac{100\eqcnc}{n_H}\Big) \Big(\frac{9.9}{1+8.9Z}\Big) ^{0.37}, 
\end{align}
where $\mathcal{N}_{\text{trans}}$ is the transition column density between \htwo\ and \hi, $F_{LW\chi}$ is the incident LW flux in units of $\chi$, and $\alpha G$ is a dimensionless parameter for the dust optical depth. Without any self-shielding in our model, the dissociating LW photons penetrate too deeply into the high-density region and too much \hi\ forms as compared to equation \eqref{eqn:ntrans}. The general trends, however, are correct. Lower densities and higher fluxes yield deeper transitions.

We use equation \eqref{eqn:ntrans} to calibrate our \htwo\ line overlap self-shielding model, and test a range of self-shielding factors to find the optimal value. The middle plot in Fig. \ref{fig:1dss} shows that a constant $S_{1\text{\htwo}}^s\sim$400 boost to the destruction of LW radiation (equation \ref{eqn:ss}) gives a realistic match to the analytical transition point, especially for lower densities. It is less accurate for higher densities and at smaller visual extinctions. Our intention for this work is full galaxy simulations where we do not resolve column densities this small and the grid cell would be effectively entirely molecular at this density. We run the test again with the temperature allowed to vary (bottom of Fig. \ref{fig:1dss}) and find the match between the analytic expression and our model to be close.

\begin{figure}
\begin{tabular}{l}
\includegraphics[width=\columnwidth]{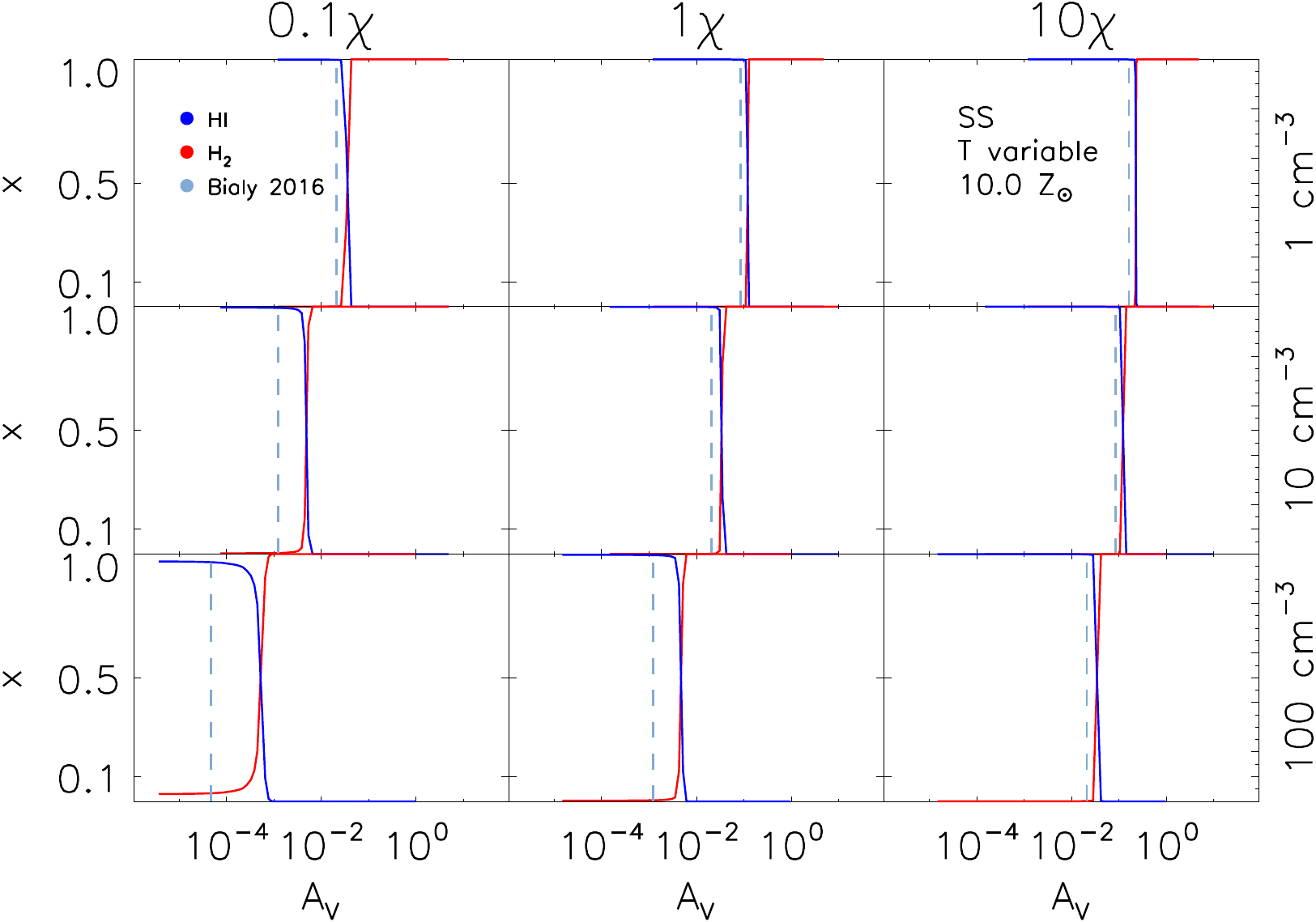}\\
\includegraphics[width=\columnwidth]{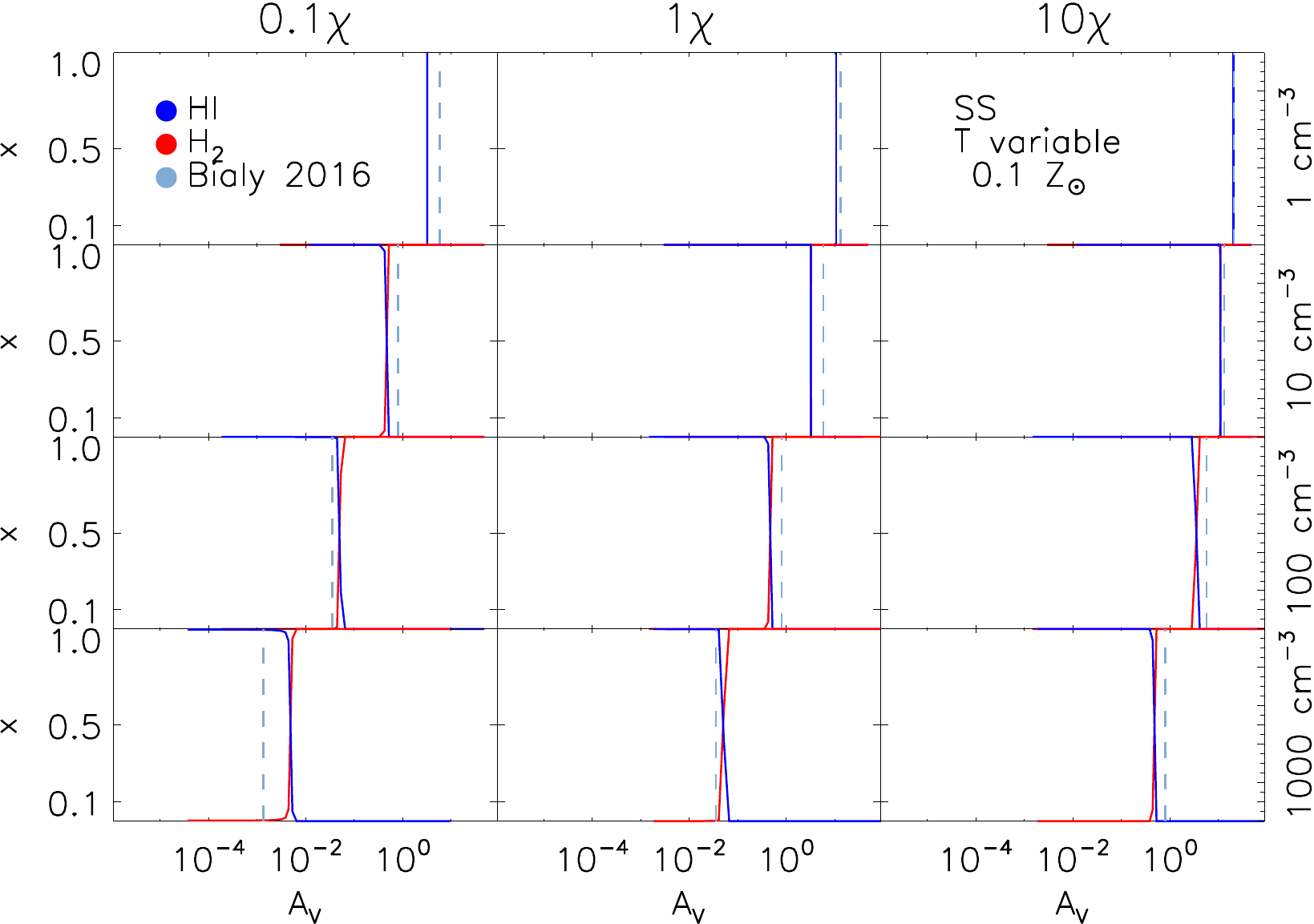}
\end{tabular}
\caption{Same as the bottom plot in Fig. \ref{fig:1dss} with self-shielding and variable temperature, but with ten times Solar metallicity (top) and a tenth of Solar metallicity (bottom).}
\label{fig:1dssmet}
\end{figure}

We explore metallicity dependence in Fig. \ref{fig:1dssmet}, by rerunning the same densities and fluxes with variable temperature and self-shielding, but with ten times and a tenth of Solar metallicity. In the higher-metallicity case, we exclude 1000 $\eqcnc$ because of the extremely small scales of the transition region. Here our transition and equation \eqref{eqn:ntrans} are close for high column density transitions, with slight underprediction of the molecular region size, and they disagree more at the lowest transition column densities. In the low-metallicity case, we tend to slightly overpredict the size of the molecular region in the high column density cases, and more closely predict the low column density transitions. In both of these extreme metallicity situations our transition point follows the correct trend where the \htwo\ region is thicker for higher metallicity and thinner for lower metallicity, and we predict the transition depth closely enough for the purposes of galaxy simulations.

\subsection{One-dimensional photodissociation regions}
\label{sec:1d}

\citet{Roellig2007} present a series of benchmark tests for 10 PDR codes, not to mimic any specific astrophysical scenario but instead as a reference by which to understand present and compare future PDR models. The codes they use are \textsc{cloudy} \citep{Ferland1998}, \textsc{costar} \citep{Kamp2000}, \textsc{htbkw} \citep{Tielens1985}, \textsc{Kosma-$\tau$} \citep{Stoerzer1996}, \textsc{Lee96} \citep{Lee1996}, \textsc{Leiden} \citep{Black1987}, \textsc{Meijerink} \citep{Meijerink2005}, \textsc{Meudon} \citep{LeBourlot1993}, \textsc{Sternberg} \citep{Sternberg1989}, and \textsc{UCL\_PDR} \citep{Taylor1993}. In these benchmark tests a plane-parallel, one-dimensional, optically thick \htwo\ slab is illuminated unidirectionally by a constant LW flux. They tested eight scenarios: $n=10^3$ and $10^{5.5} \eqcnc$, $F_{LW}=10$ and $10^5 \chi$, with temperature fixed at 50 K and variable. To date, the only other galaxy code to run this comparison is \citet{Baczynski2015}, where the hydrogen tests are favourable.

We compare our model to these benchmark tests for the $n=10^3\eqcnc$ cases. We do not test the $n=10^5\eqcnc$ cases because of the extremely high resolution required to resolve the thin \hi-\htwo\ transition layer, which our code is not specialized to do. At such high densities, the region is essentially entirely molecular. We use Solar metallicity for these tests. The boxsize is 10 pc with the AMR grid resolution between 256$^1$ and 16384$^1$ cells. The cell column density of the refined regions are $2\times 10^{18}$ cm$^{-2}$. Cosmic ray ionization and heating are present.

\begin{figure*}
\includegraphics[width=0.85\textwidth]{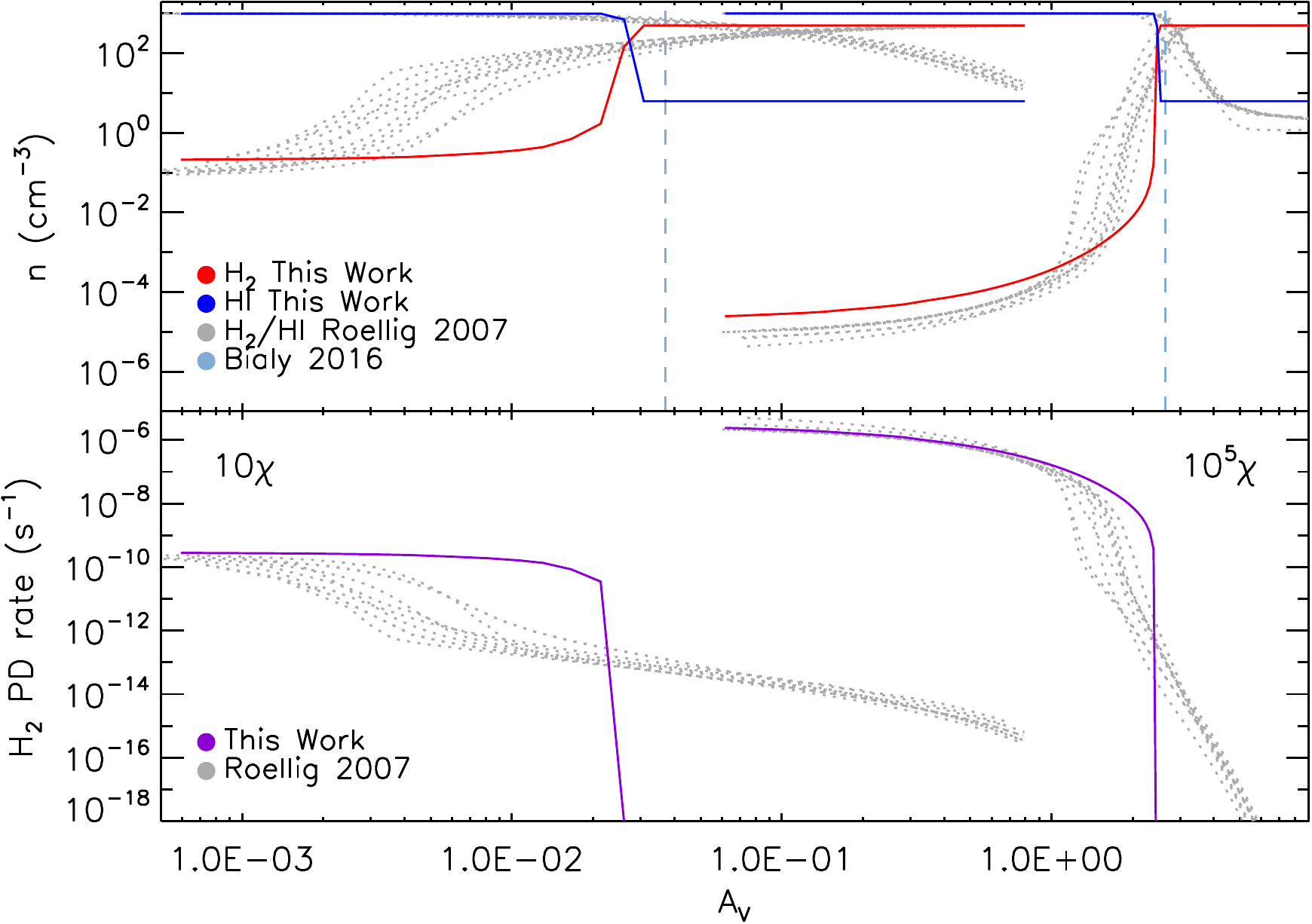}
\caption{Hydrogen fraction number density (top) and photodissociation rate (bottom) versus visual extinction of a one-dimensional region for fixed density $10^3$ cm$^{-3}$ and fluxes 10 and 10$^5 \chi$, compared to \citeauthor{Roellig2007}'s (\citeyear{Roellig2007}) PDR simulations and \citeauthor{Bialy2017}'s (\citeyear{Bialy2017}) transition. The temperature is fixed.}
\label{fig:1dtfix}
\end{figure*}

Fig. \ref{fig:1dtfix} shows the high-density region profiles of the number density of \hi\ and \htwo\ and the LW photodissociation rate, for the $F_{LW}=10,$ and $10^5 \chi$  cases and a fixed $T=50$K. The most striking feature is that the transition between atomic and molecular happens more abruptly in our model as compared to that of the \citet{Roellig2007} benchmark tests. In both the density and the photodissociation rate profiles we do not reproduce the gradual transition. This is expected due to a difference in how we handle \htwo\ self-shielding. The traditional Draine \citep{Draine1996} and Draine-inspired functions follow a power law, while our constant factor leads to an exponential cut off. Nevertheless, this is incidental on the scales we will use for our galaxy simulations. We mainly seek to reproduce the transition depth, which is close enough for our purposes. We also draw \citeauthor{Bialy2017}'s (\citeyear{Bialy2017}) transition point for reference. In the $10 \chi$ case, it is left to the scatter of the transition as predicted by the \citet{Roellig2007} tests, as is our transition. For $10^5 \chi$, our model, the \citet{Roellig2007} tests, and \citeauthor{Bialy2017}'s (\citeyear{Bialy2017}) transition are all in agreement. Our residual \hi\ in the \htwo\ region, caused by cosmic rays, is also close to that of \citet{Roellig2007}.

\begin{figure*}
\includegraphics[width=0.85\textwidth]{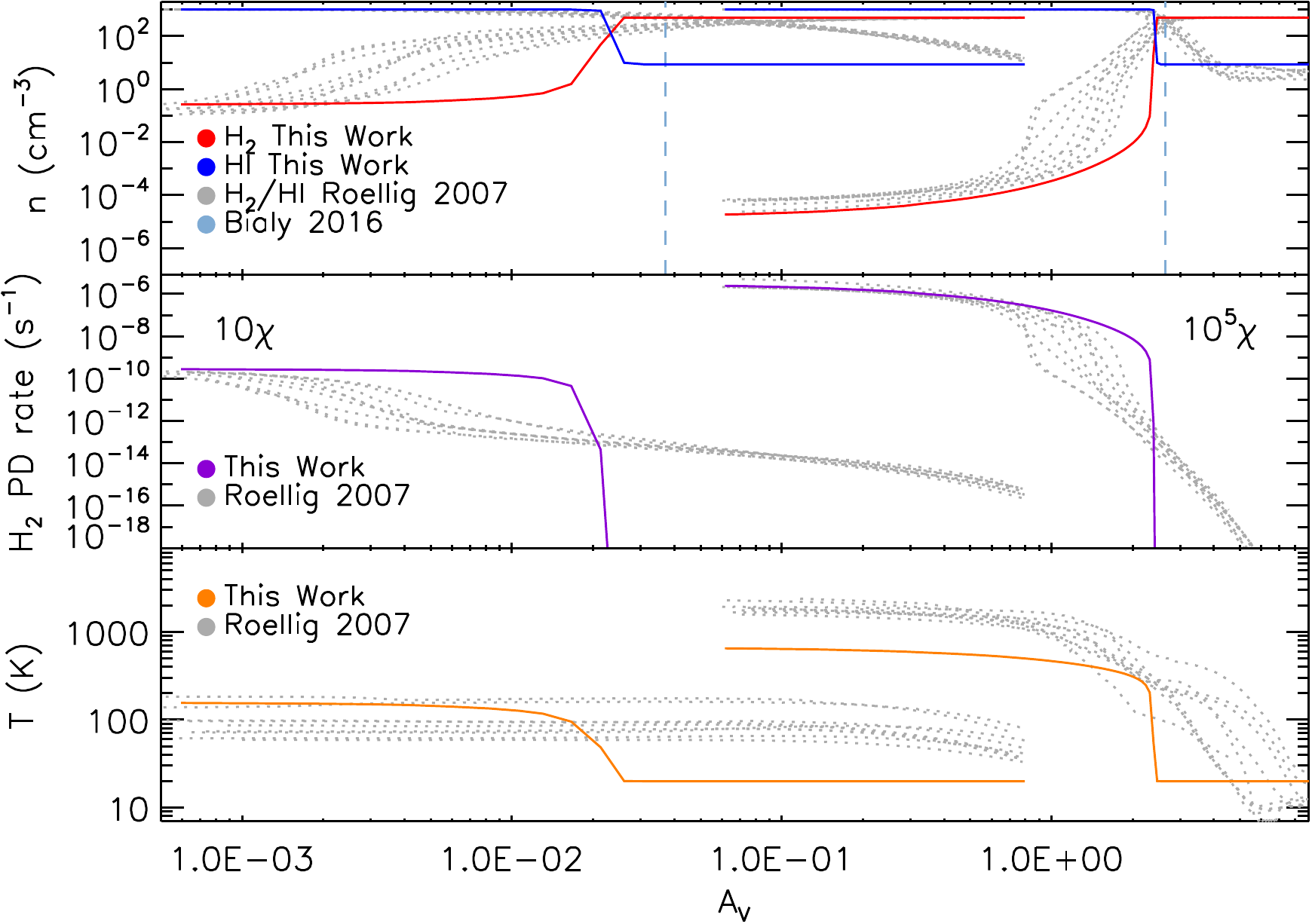}
\caption{Hydrogen fraction number density (top), photodissociation rate (middle), and temperature (bottom) versus visual extinction of a one-dimensional region for fixed density $10^3$ cm$^{-3}$ and fluxes 10 and 10$^5 \chi$, compared to \citet{Roellig2007}'s PDR simulations and \citet{Bialy2017}'s transition. The temperature is variable.}
\label{fig:1dtvar}
\end{figure*}

Fig. \ref{fig:1dtvar} gives the same profiles as Fig. \ref{fig:1dtfix}, except with the addition of the temperature profile for the variable-temperature case. Here our transition between atomic and molecular happens much more closely to that of the \citet{Roellig2007} benchmark tests, while the abruptness in transition shape remains. The cooling of the high-density region is also more abrupt in our model, following our exponential model for self-shielding as explained above. This highlights how \htwo\ dominates the cooling process, as seen also in the single cells in Section \ref{sec:singlecell}. In our high flux case, $10^5 \chi$, the atomic region is cooler compared to those of the benchmark tests, similar to how \citet{Baczynski2015} find that their code is cooler in this region, and these discrepancies are likely due to the different cooling models.

PDR codes are overwhelmingly one-dimensional. However, two codes built for three dimensions compare their models to the \citet{Roellig2007} benchmark. \textsc{3d-pdr} \citep{Bisbas2012} takes the chemistry of \textsc{ucl\_pdr} and ray tracing applied to a three-dimensional cloud of arbitrary density distribution. \textsc{km2} \citep{Motoyama2015} is a hybrid hydrodynamical and chemical code. These codes are tested first in one dimension to directly compare to \citet{Roellig2007}'s tests, and test a sphere or radius 5.15 pc hit by plane-parallel radiation, in a setup similar to the benchmark test for $n=10^3\eqcnc$, $F_{LW}=10\chi$, and variable temperature. They find that their spheres generally agree with the benchmark from one dimension. Our code works in higher dimensions \citep{Rosdahl2013} and accordingly the one-dimensional PDR tests we show in this section are sufficient.

\subsection{Str{\"o}mgren spheres in a molecular medium}

A Str{\"o}mgren sphere \citep{Stromgren1939} describes the growth of an ionization front around a radiation source embedded in a neutral medium of hydrogen density $n_H$, assuming an infinite speed of light. In three dimensions, the radius of the ionization front increases with time as
\begin{equation}
r_{I}(t)=r_{S\text{\hi}}(1-e^{-t/t_{rec\text{\hi}}})^{1/3},
\label{eqn:stromhi}
\end{equation}
where $r_{S\text{\hi}}$ is the Str{\"o}mgren radius at which recombination balances ionization and $t_{rec\text{\hi}}$ is the recombination time, given by
\begin{equation}
r_{S\text{\hi}} = \Big(\frac{3 \dot{N}_{\text{\hi}}^{\text{er}}}{4\pi\alpha_{\text{\hi}} n_H^2}\Big)^{1/3},
\label{eqn:rshi}
\end{equation}
\begin{equation}
t_{rec\text{\hi}}=(n_H\alpha_{\text{\hi}})^{-1}.
\label{eqn:trec}
\end{equation}
$\dot{N}_{\text{\hi}}^{\text{er}}$ is the ionizing photon emission rate and $\alpha_{\text{\hi}}$ is the recombination rate of \hi.

\citet{Iliev2006} provide two tests for radiative transfer codes using the Str{\"o}mgren sphere framework. A source of \hi-ionizing radiation emits at a rate of $\dot{N}_{\text{\hi}}^{\text{er}}=5\times10^{48}$ photons s$^{-1}$ in a homogeneous, neutral medium of density $n_H=10^{-3}$ cm$^{-3}$, with a resolution of 128$^3$ cells. The evolution of the resulting ionization front is then compared between codes and against the analytic solution. First the temperature is fixed at 10$^4$ K and in the second test the temperature is allowed to vary. \citet{Rosdahl2013} successfully compares these two tests to both the analytic solution and other codes.

We extend these tests to involve \htwo. Equivalent to the Str{\"o}mgren sphere's equation \eqref{eqn:stromhi} and assuming an infinite light speed, the radius of the \htwo\ dissociation front should grow as
\begin{equation}
r_{D}(t)=r_{S\text{\htwo}}(1-e^{-t/t_{rec\text{\htwo}}})^{1/3},
\label{eqn:stromh2}
\end{equation}
where $r_{S\text{\htwo}}$ is a molecular Str{\"o}mgren radius given by,
\begin{equation}
r_{S\text{\htwo}} = \Big(r_{S\text{\hi}}^3+\frac{3\dot{N}_{\text{\htwo}}^{\text{er}}}{4\pi\alpha_{\text{\htwo}} n_H^2}\Big)^{1/3}.
\label{eqn:rsh2}
\end{equation}
$\dot{N}_{\text{\htwo}}^{\text{er}}$ is the dissociating LW photon emission rate. The formula for the \htwo\ recombination time, $t_{rec\text{\htwo}}$, is the same as for \hi\ (equation \ref{eqn:trec}), only with the corresponding $\alpha_{\text{\htwo}}$ as formation rate. These equations for the \htwo\ sphere are analogous to the \hi\ sphere (equation \ref{eqn:rshi}) and we note that in equation \eqref{eqn:rsh2}, we add the \hi\ radius to the \htwo\ radius because the \htwo\ sphere is expected to grow from the \hi\ sphere.

For our simulations, we keep the density at $n_H=10^{-3}$ cm$^{3}$ and begin with a fully molecular medium. The source is a supposed $4.3\times10^4$ K O star with radius 10 R$_{\odot}$, which yields emission rates $\dot{N}_{\text{\htwo}}^{\text{er}}=3\times10^{48}$ and $\dot{N}_{\text{\hi}}^{\text{er}}=5\times10^{48}$ photons s$^{-1}$. The \htwo\ dissociation cross-section is the same as in equation \eqref{eqn:h2diss_cs} and the ionization cross-sections are averaged over a $4.3\times10^4$ K black body: $\sigma^{N}_{2\text{\htwo}}=3.6\times 10^{-18} \text{cm}^2$ and $\sigma^{N}_{2\text{\hi}}=5.0\times 10^{-18} \text{cm}^2$.
For the fixed gas temperature test we use $3.56\times10^3$ K at which the equilibrium concentration is half molecular and half neutral, different from the initial condition. This is a little lower than the temperature for the \citet{Iliev2006} test to allow for the existence of molecular gas. We use Solar metallicity, a boxsize of 10 kpc, resolution 128$^3$ cells, the GLF flux function, and run it for 500 Myr. For this set up, $r_{S\text{\htwo}}=295$ kpc, $r_{S\text{\hi}}=4.10$ kpc, $t_{rec\text{\htwo}}=3.33\times10^7$ Myr, and $t_{rec\text{\hi}}=53.7$ Myr. For both the fixed-temperature and the variable-temperature tests we use two situations: without \htwo\ self-shielding, and fully shielded. The OTSA is used, and the light speed fraction is set to 10$^{-2}$ as in \citet{Rosdahl2013}. We neglect cosmic rays here to keep the comparison as similar as possible to the original tests.

\citet{Shapiro2006} give a relativistic expression for the \hi\ ionization front expansion that takes a non-infinite speed of light into account:
\begin{align}
\label{eqn:shapiro}
w&=qy-\mathrm{ln}(1-y^3),\\
w&\equiv t/t_{rec\text{\hi}}, \\
y&\equiv r_I/r_{S\text{\hi}}, \\
q&\equiv r_{S\text{\hi}}/(c_{\text{r}} t_{rec\text{\hi}}).
\end{align}
For a more realistic comparison of our numeric simulation to this formulation, we will use the reduced speed of light, $c_{\text{r}}$, instead of the full speed of light. Deriving an equivalent formula for the \htwo\ dissociation front is beyond the scope of this paper.

\begin{figure}
\begin{tabular}{l}
\includegraphics[width=\columnwidth]{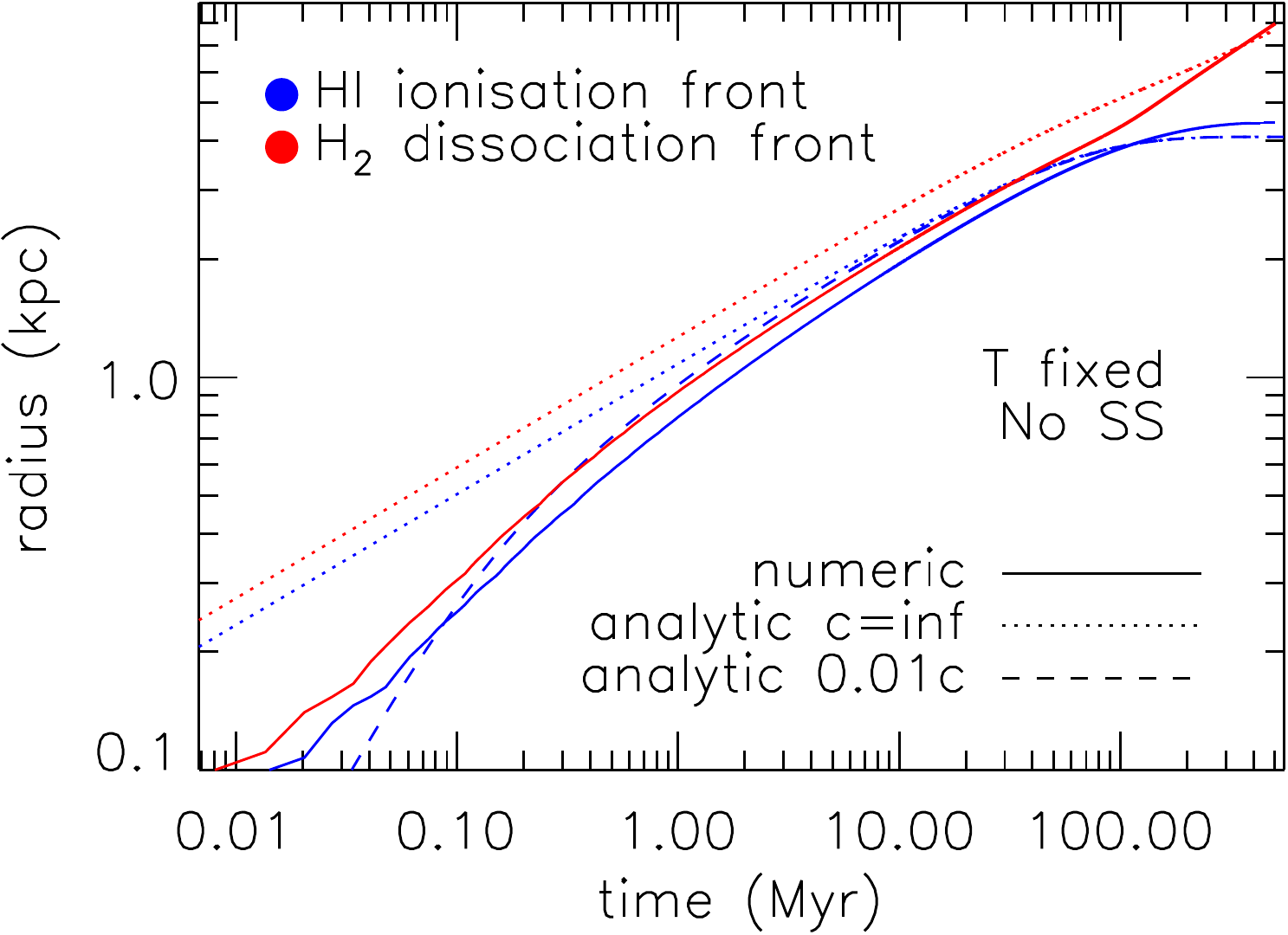} \\
\includegraphics[width=\columnwidth]{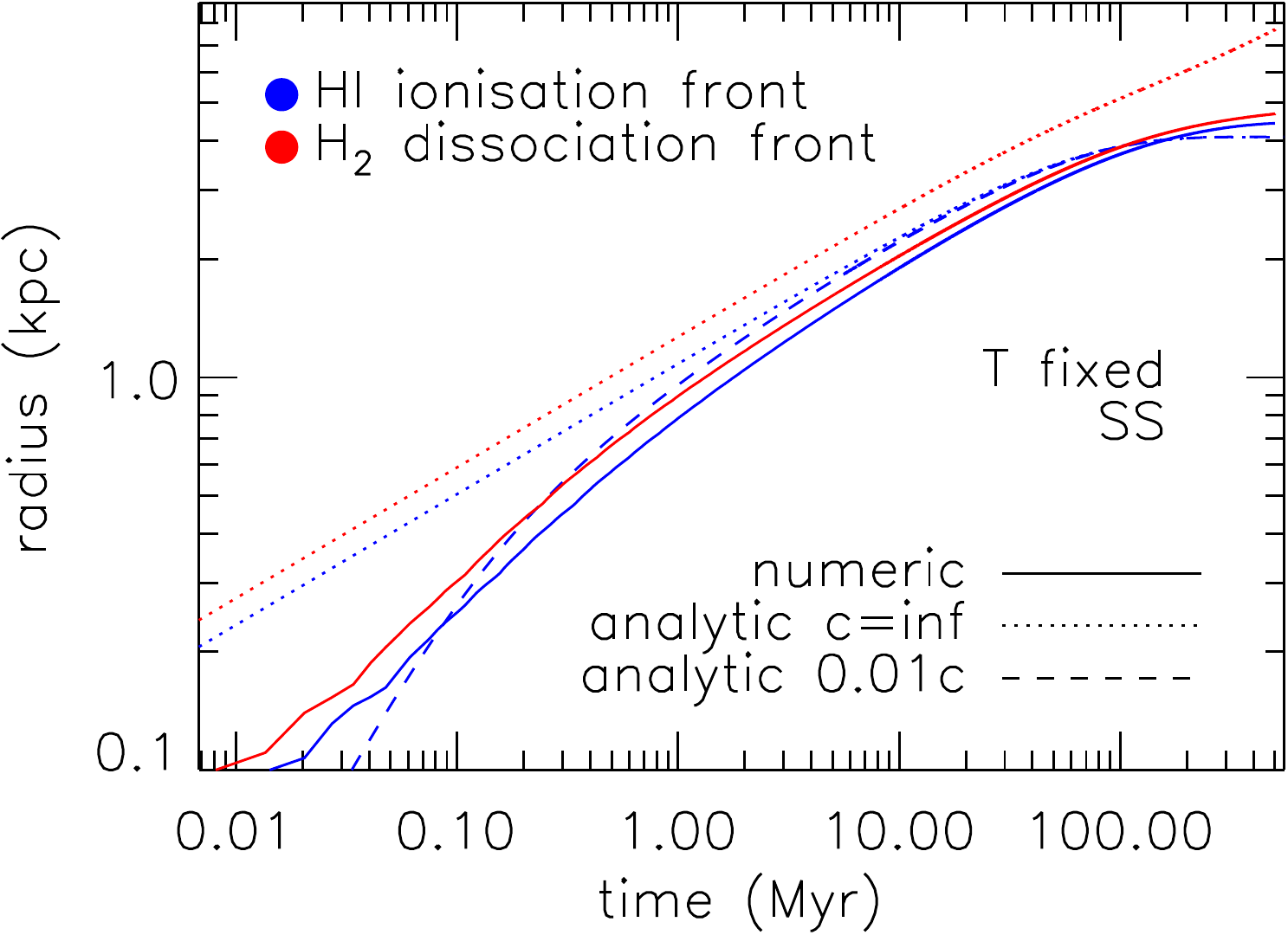}
\end{tabular}
\caption{The evolution of \hi\ ionization and \htwo\ dissociation fronts up to 500 Myr for a Str{\"o}mgren-like scenario with fixed temperature and 10$^{-3}\eqcnc$. Top: without self-shielding of \htwo. Bottom: with \htwo\ self-shielding. The boxsize is 10 kpc. Solid lines follow our simulations. The dotted lines follow equations \eqref{eqn:stromhi} and \eqref{eqn:stromh2}. The dashed line follows the reduced speed of light equation given in \citet{Shapiro2006}.}
\label{fig:strom_ifront_tfix}
\end{figure}

Fig. \ref{fig:strom_ifront_tfix} gives the evolution of the \htwo\ dissociation and \hi\ ionization fronts without and with shielding for the fixed-temperature scenario. Their evolution is compared to the analytic equations \eqref{eqn:stromhi} and \eqref{eqn:stromh2} for the infinite speed of light and equation \eqref{eqn:shapiro} for the \textbf{finite} speed of light. Before t$_{rec\text{\hi}}$, both the shielded and non-shielded cases grow similarly. They grow much more slowly compared to their analytic components because of the reduced speed of light, and more closely with the \citet{Shapiro2006} relativistic expression. This is also quite similar to what \citet{Rosdahl2013} found, where the analytic front is ahead of the numeric front by about 5 per cent because the numeric front evolves more gradually than the step-wise analytic front. 

Once t$_{rec\text{\hi}}$ has passed, the numeric \hi\ fronts catch up to the analytic expressions and level off at a radius of about 5 kpc close to the calculated Str{\"o}mgren radius of about 4 kpc. Concerning the \htwo\ front, the unshielded and shielded cases differ after t$_{rec\text{\hi}}$. In the unshielded case, the \htwo\ front continues to grow and reaches 8 kpc at 500 Myr, the simulation end time. It would continue to grow, given that $t_{rec\text{\htwo}}\sim 10^7$ Myr, but in reality this is much longer than the age of Universe. The shielded case, on the other hand, demonstrates the importance of \htwo\ self-shielding. Here the \htwo\ front levels off much like the \hi\ front, extending only slightly beyond it at around 5 kpc. This is expected because our analytic expressions do not take shielding into account.

\begin{figure*}
\begin{tabular}{l}
\includegraphics[width=0.85\textwidth]{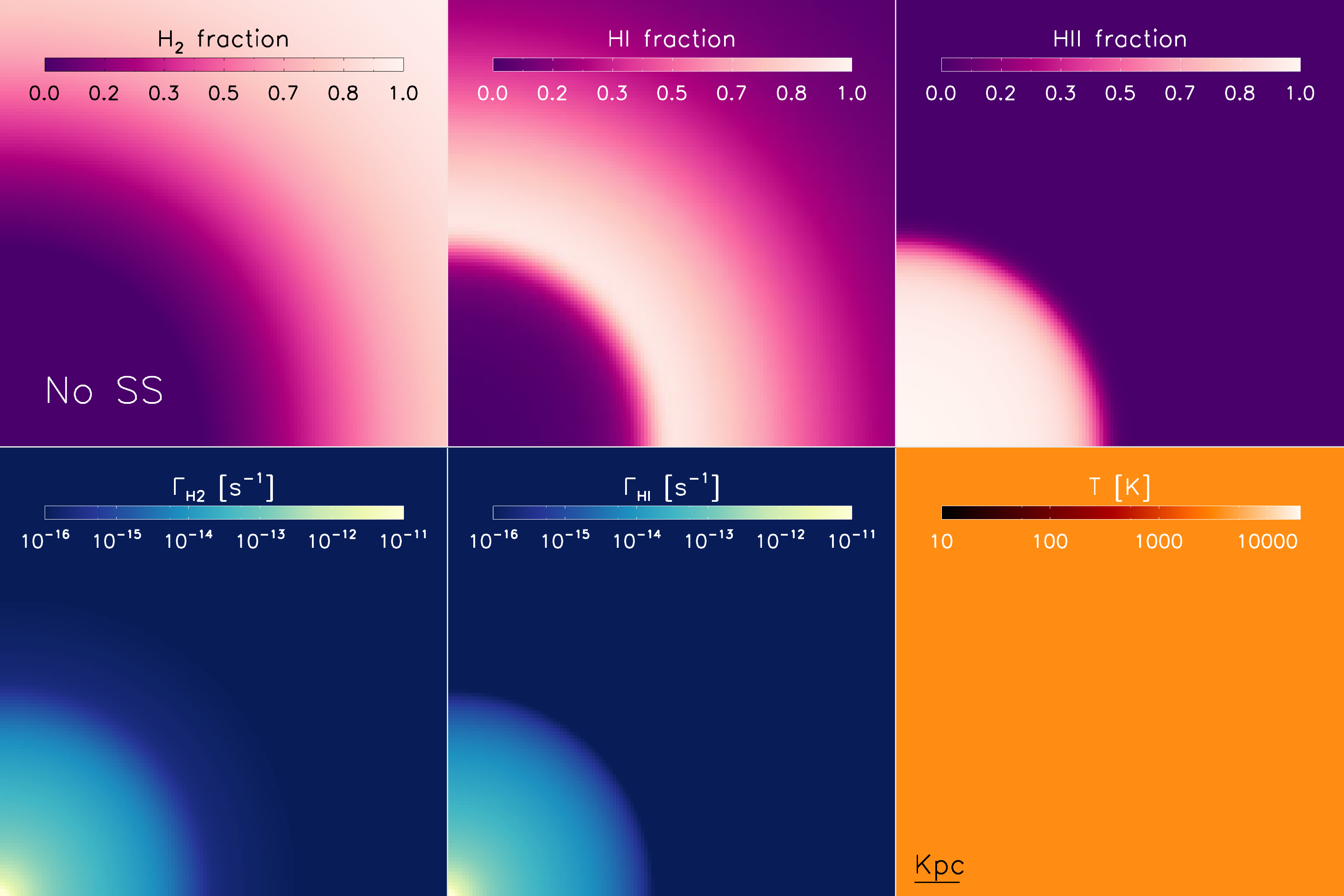}\\
\includegraphics[width=0.85\textwidth]{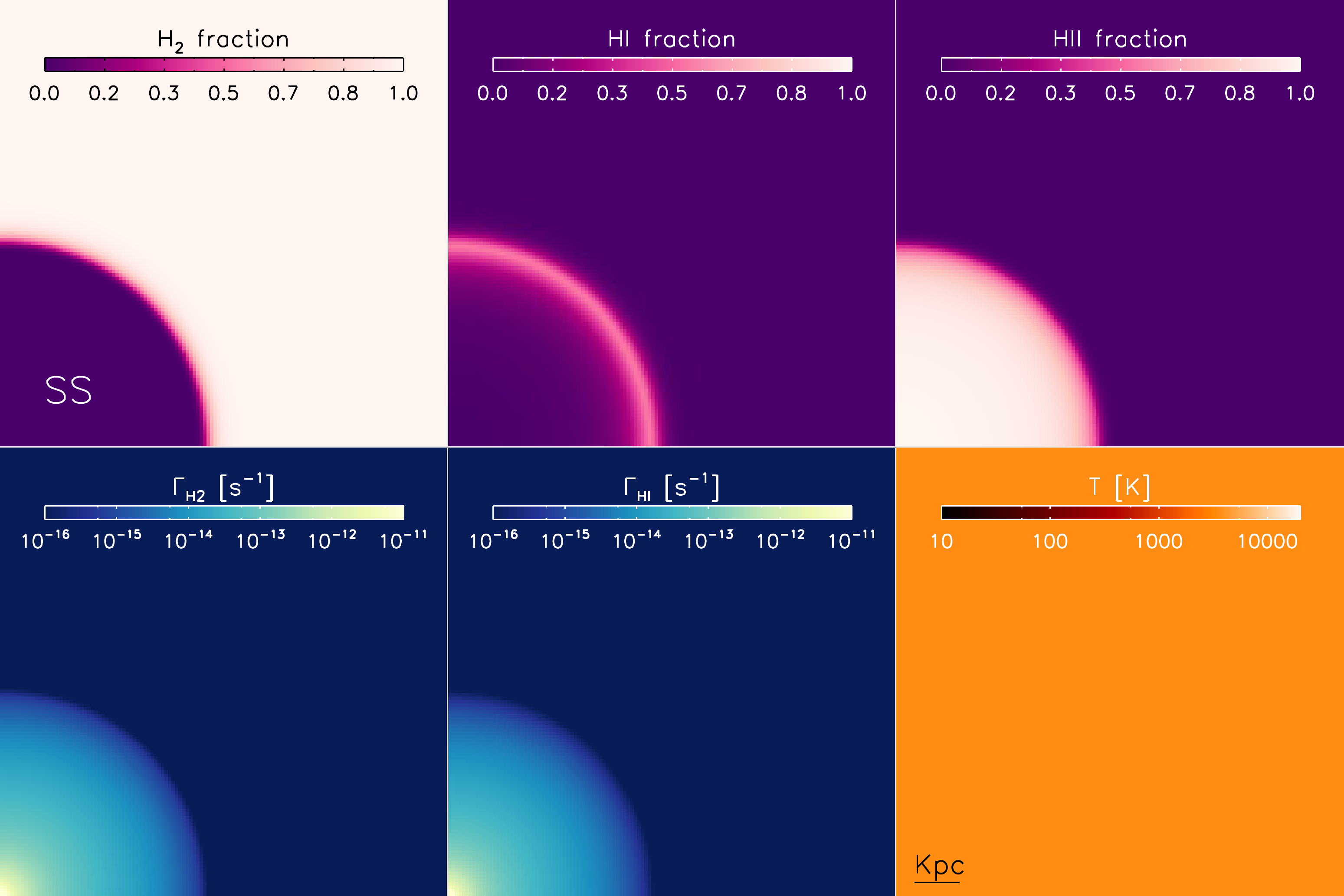}
\end{tabular}
\caption{Central slices for our Str{\"o}mgren-like scenario at 500 Myr, with fixed temperature and boxsize 10 kpc. Contours are given for \htwo, \hi, and \hii\ fractions, \htwo\ and \hi\ photodissociation rates, and temperature. Top two rows: unshielded case. Bottom two rows: shielded case.}
\label{fig:strom_cont_tfix}
\end{figure*}

Fig. \ref{fig:strom_cont_tfix} shows the hydrogen fractions and radiation maps at 500 Myr, for the unshielded and shielded cases with fixed temperature. In both cases, \hi\ stops the ionizing photons and the \hii\ region ends sharply. In the unshielded case, the dissociating LW photons extend much further into the \htwo\ layer. In the shielded case, the \htwo\ is able to completely block the LW photons and maintain a pure molecular layer.

\begin{figure}
\begin{tabular}{l}
\includegraphics[width=\columnwidth]{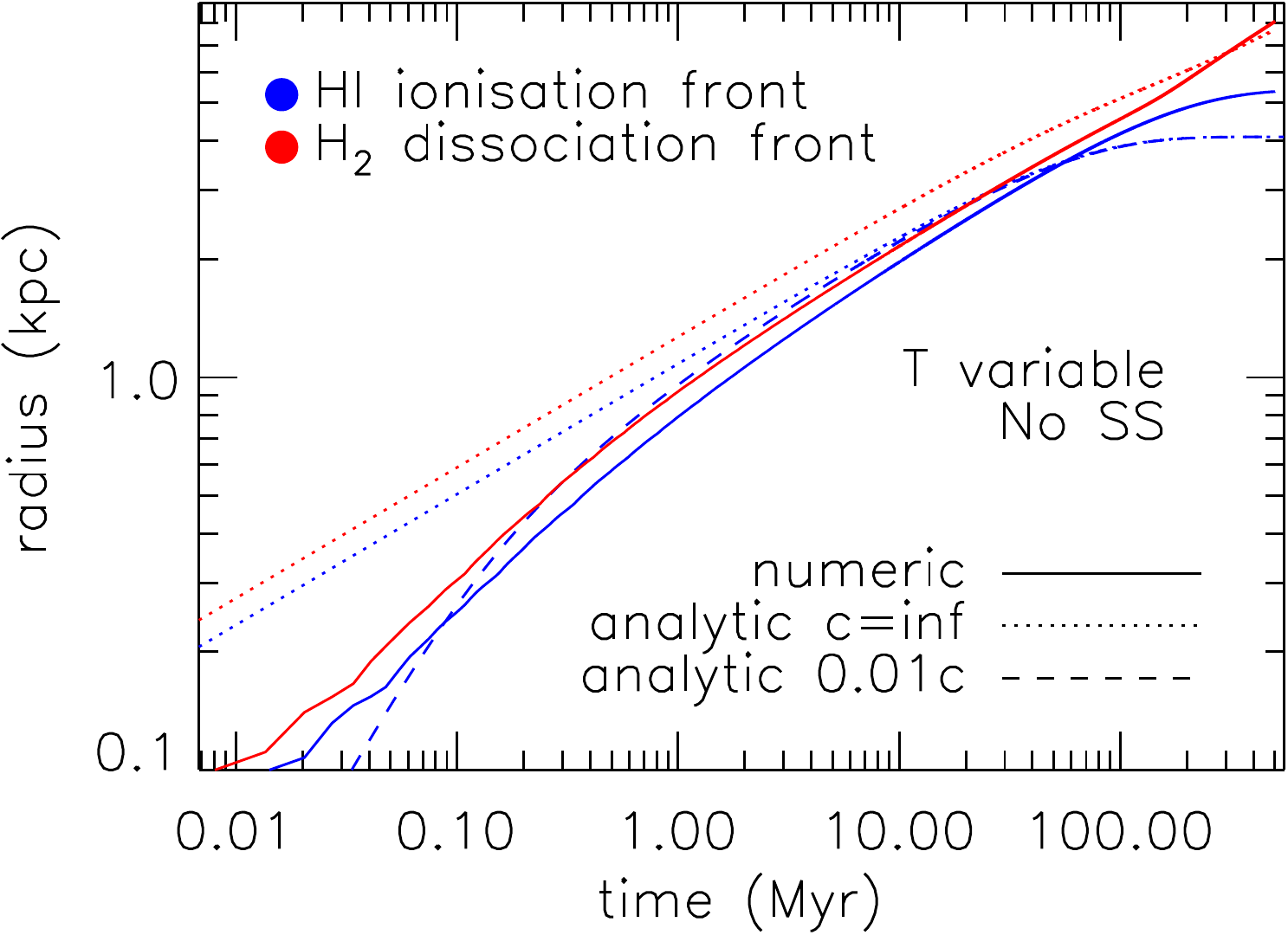} \\
\includegraphics[width=\columnwidth]{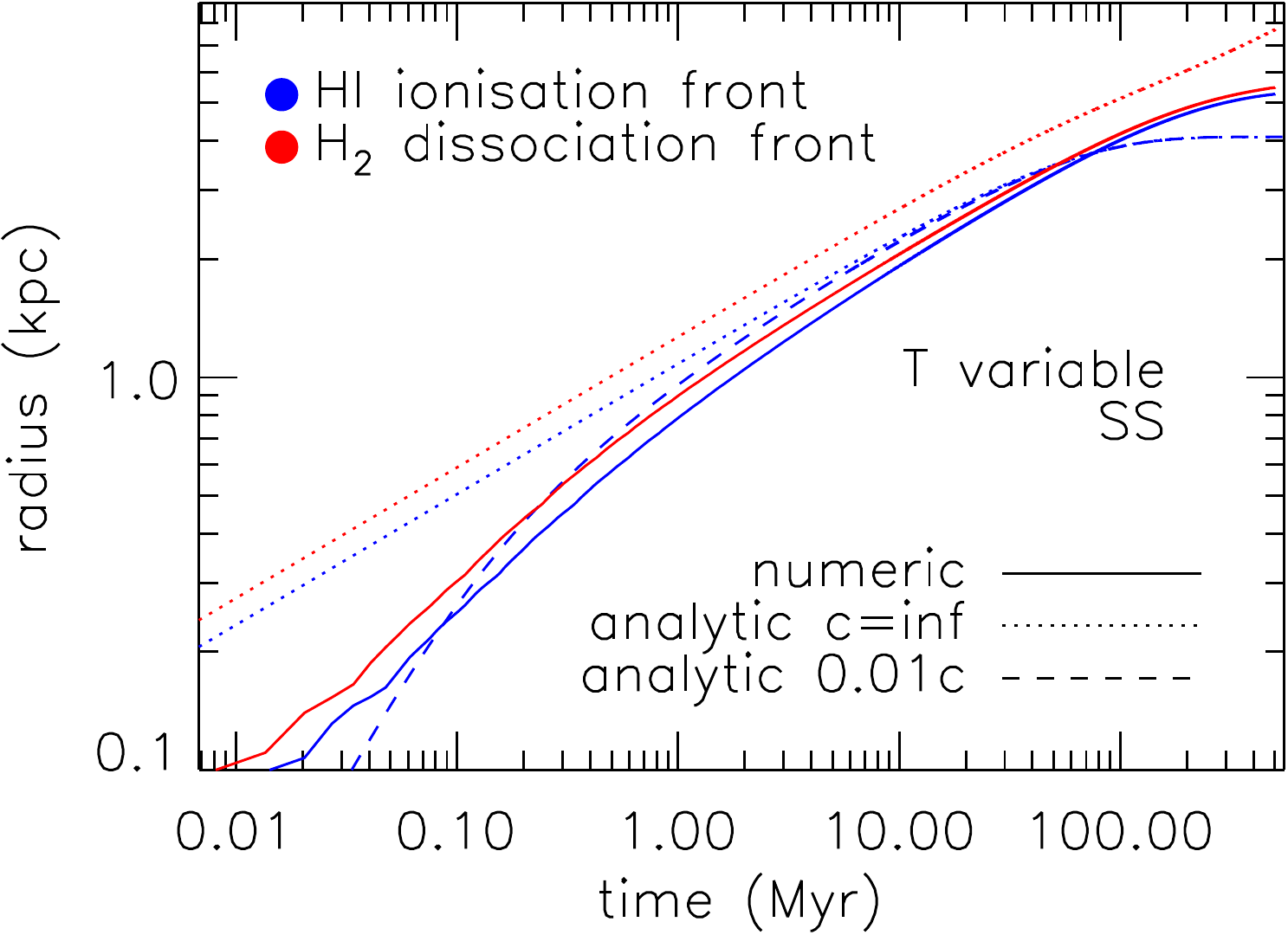}
\end{tabular}
\caption{The evolution of \hi\ ionization and \htwo\ dissociation fronts up to 500 Myr for a Str{\"o}mgren-like scenario with variable temperature and density 10$^{-3}\eqcnc$. Top: without self-shielding of \htwo. Bottom: with \htwo\ self-shielding. The boxsize is 10 kpc. Solid lines follow our simulations. The dotted lines follow equations \eqref{eqn:stromhi} and \eqref{eqn:stromh2} for fixed $T=3.56\times10^3$ K. The dashed line follows the reduced speed of light equation given in \citet{Shapiro2006} for fixed $T=3.56\times10^3$ K.}
\label{fig:strom_ifront_tvar}
\end{figure}

Next, Fig. \ref{fig:strom_ifront_tvar} gives the \htwo\ and \hi\ fronts for the variable-temperature scenario, both unshielded and shielded. The analytic expressions from Fig. \ref{fig:strom_ifront_tfix}  are left on for reference, but they are less relevant here because of the changes in the formation rates, and hence recombination times and Str{\"o}mgren radii. With variable temperature, the growth of the fronts are similar to the fixed-temperature case before t$_{rec\text{\hi}}$. After one recombination time, the \hi\ fronts level off to a radius  larger than in the fixed-temperature case. Also, as in the fixed-temperature case, the unshielded \htwo\ front continues to grow towards the edge of the box, while the shielded \htwo\ front follows the evolution of the \hi\ front at a slightly larger radius. 

\begin{figure*}
\begin{tabular}{l}
\includegraphics[width=0.85\textwidth]{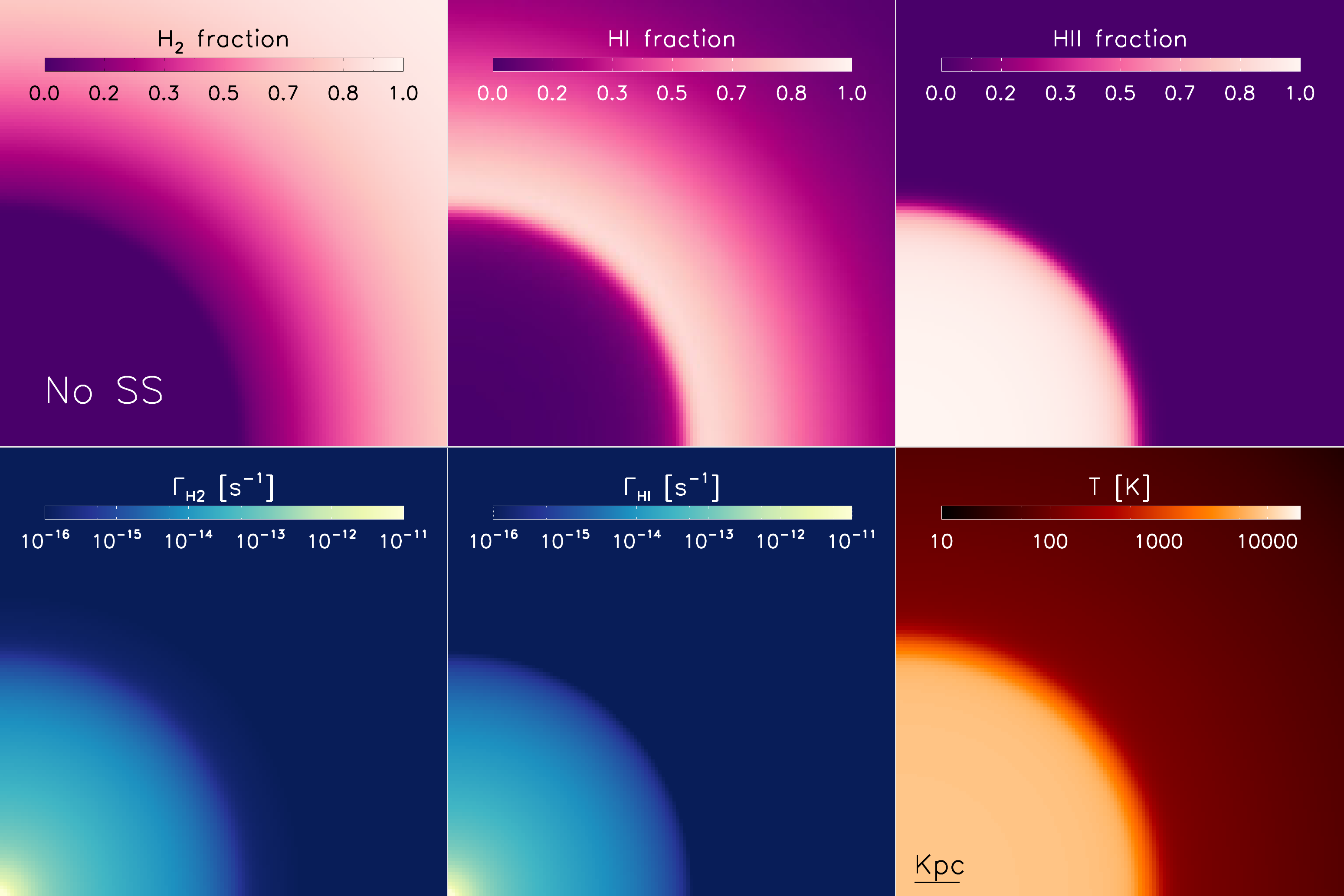}\\
\includegraphics[width=0.85\textwidth]{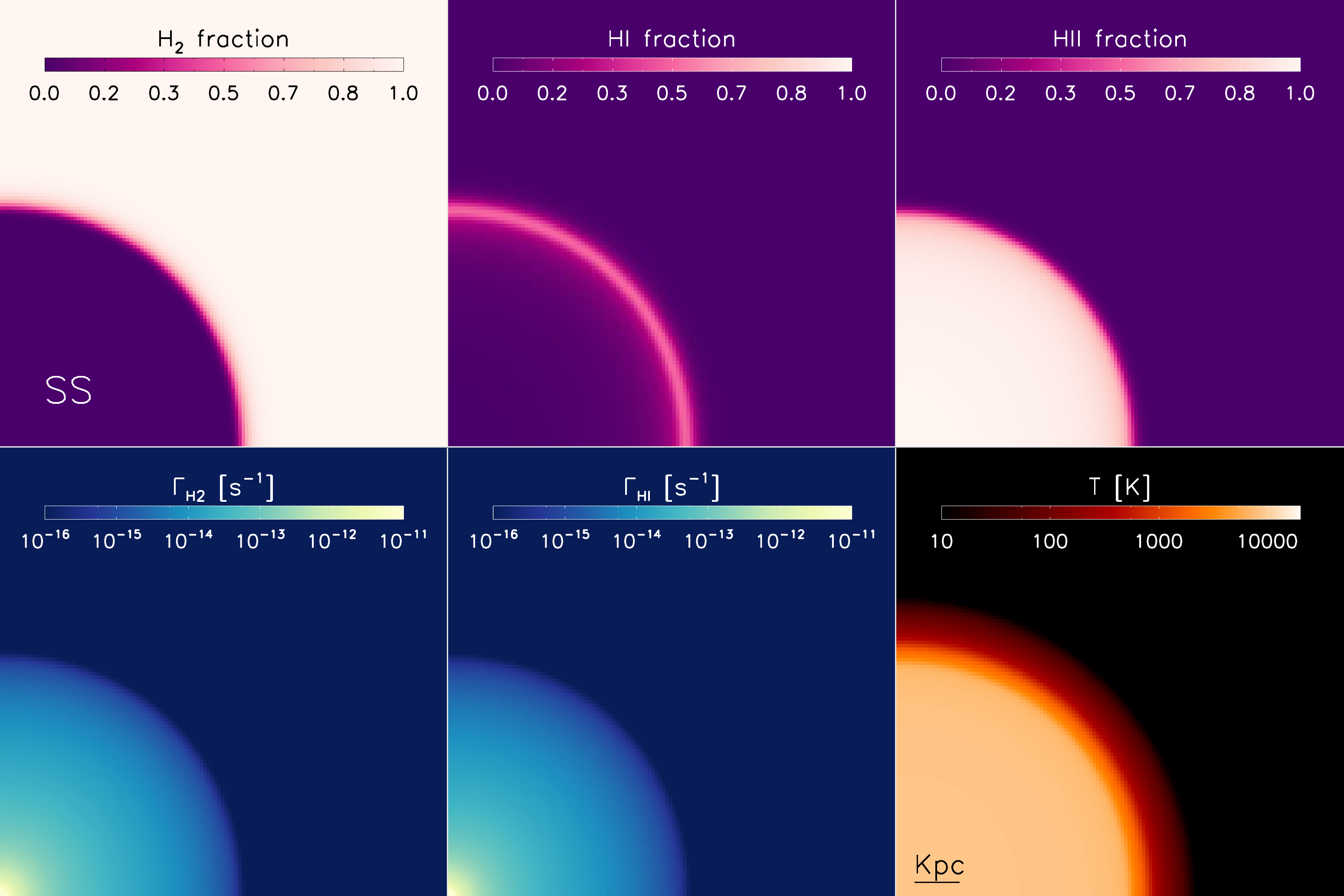}
\end{tabular}
\caption{Central slices for our Str{\"o}mgren-like scenario at 500 Myr, with variable temperature and boxsize 10 kpc. Contours are given for \htwo, \hi, and \hii\ fractions, \htwo\ and \hi\ photodissociation rates, and temperature. Top two rows: unshielded case. Bottom two rows: shielded case.}
\label{fig:strom_cont_tvar}
\end{figure*}

In Fig. \ref{fig:strom_cont_tvar}, the maps of hydrogen fractions and their radiation are similar to their fixed-temperature counterparts. The point of interest is in comparing the temperature maps for the unshielded and shielded cases. When \htwo\ is shielded, the molecular region cools to the $\sim$10 K floor. Unshielded, the molecular region still has atomic content and cools to only $\sim$100 K. This is reminiscent of our single cell tests (Section \ref{sec:singlecell}), where in presence of a UV background the lower-density cells were unable to cool to $\sim$10 K. Self-shielding is critical to \htwo\ formation and molecular cooling. \citet{Baczynski2015} presents similar molecular Stro{\"o}mgren tests, though with hydrodynamics and a fixed recombination rate, and like us observe the thin atomic layer between \htwo\ and \hii.

Our adaptation of the Str{\"o}mgren sphere to a situation involving both \htwo\ and \hi\ is realistic. Our numeric results are in line with the analytic framework, and where they differ it is explained. On the shorter time-scales our fronts grow more slowly than those of the analytic framework, and this is caused by the reduced speed of light for faster computation. If we do simulations where we are interested in shorter time-scales then we should use the full speed of light. However, once we reach time-scales of tens of Myr and higher, our simulations grow as the analytical functions. It is these longer time-scales that are of interest to our galactic application of this methodology. 

\section{Summary and future}
\label{sec:conc}

In this paper we present our molecular addition to \textsc{Ramses-RT}, an AMR hydrodynamical code with radiative transfer. We follow the non-equilibrium evolution of molecular, atomic, and ionized hydrogen coupled to the radiative transfer of the dissociating Lyman-Werner and ionizing photons. Our moment-based radiative transfer uses the Eddington tensor approximation for closure. Because this method is purely local, we gain tremendous computational efficiency independent of source number. A semi-implicit method advances our thermal chemistry rate equations in time, and species fractional abundances are fully coupled to temperature, radiation, and hydrodynamics. The chemical processes we include for \hi\ are recombination, destruction by electron collision, and photoionization; for \htwo\ we include formation catalysed by dust grains, primordial gas phase formation in the absence of metals, formation by three-body collisions, collisional destruction with atomic hydrogen and itself, and photodestruction by dissociating LW photons and higher-energy ionizing photons. We have also added cosmic ray ionization of \htwo, \hi, and \hei.

We capitalize on our moment-based radiative transfer to introduce a new method of modelling \htwo\ self-shielding against LW photons. We boost the destruction of LW photons that dissociate \htwo\ by a constant factor to incorporate the fact that only a fraction of LW photon absorption leads to \htwo\ dissociation. As the LW photons continue to travel through gas cells rich in \htwo\ across many time-steps, their repeated destruction mimics a column density. This differs from works by other authors where \htwo\ self-shielding is implemented by converting a volume density to a column density and decreasing \htwo\ destruction. 

A suite of tests demonstrate the robustness of our method across an array of situations.

\textbf{Single cells}: Our single-cell tests evolve the hydrogen chemistry in zero dimensions, for a grid of initial temperatures, fixed densities, and initial atomic/ionized fractions. The four scenarios are as follows: with fixed or variable temperatures and with or without a UV background flux. In the fixed-temperature cases, the cells evolve to the expected equilibrium states given enough time. Around 10$^4$ K the final state is entirely atomic, while at higher temperatures the final state is fully ionized and at lower temperatures it is fully molecular. With a UV background, the final state is also dependent on density and higher-density cells give increasingly molecular final states. With evolving temperature, the cells cool down to the expected $\sim$10 K floor. Cooling occurs faster with increasing density and decreasing initial ionization fraction. In the presence of a UV background, lower-density cells are unable to cool down to this floor, and the final temperature is dependent on cell density.

\textbf{Self-shielding calibration}: We calibrate our self-shielding model with one-dimensional simulations. A constant flux of LW photons hits a high-density \htwo\ region, and we repeat this for a grid of constant densities and LW fluxes. In each high-density region, the photons dissociate the \htwo\ into \hi\ until the photons are all destroyed by dissociation, leaving an \htwo\ core. We compare our \hi-\htwo\ transition depth to the analytic prediction by \citet{Bialy2017}, and without self-shielding the photons penetrate the \htwo\ region too deeply. We experiment with constants by which to boost the LW photodestruction and find one factor that reproduces the analytic results satisfactorily for each flux, density, and metallicity. Our method works for both fixed- and variable-temperature scenarios.

\textbf{PDR code comparison}: We compare the results from our code to the one-dimensional \citet{Roellig2007} benchmark tests, which comprise 10 separate PDR codes. For a high-density region of $10^3 \eqcnc$, the four scenarios we test are fluxes $10$ or $10^5 \chi$, and the temperature constant at 50 K or variable. Our transition depth between \htwo\ and \hi\ is accurate. However, we are unable to reproduce the exact PDR transition shape. The photodissociation rate and temperature profiles follow a similar exponential trend. This is because the one-dimensional PDR codes are able to use a column density based power law for their \htwo\ self-shielding, while we use a local exponential form. Because of this, the small-scale physics of our transition region are inexact. However, because we will be applying the code to large-scale simulations the PDR curve is unimportant to us. The important quantity to reproduce is a transition zone, which we do successfully. 

\textbf{Str{\"o}mgren sphere}: The Str{\"o}mgren sphere models the growth of an OB star's ionization front. Traditionally, this test is done in a neutral medium but we expand it to a molecular medium. Analytical expressions predict the growth of these ionization and dissociation fronts. We compare our numeric results to the analytical expressions for four scenarios: with temperature fixed and variable and with and without self-shielding. An \hii\ sphere encapsulates the source, while an \hi\ shell separates it from the outer \htwo\ medium. The \hi\ front grows in line with expectations for numeric Str{\"o}mgren spheres. Our \htwo\ front grows at a speed similar to that of the \hi\ front up until the \hi\ recombination time. After this, the presence of self-shielding determines the  \htwo\ front's growth. Without self-shielding the \htwo\ front continues to grow, while with self-shielding the growth slows in step with the \hi\ front. When we vary the temperature, the molecular region cools to our $\sim$10 K floor with self-shielding, while without self-shielding the gas is unable to cool so low.

The importance of \htwo\ self-shielding is manifest in our simulations. Without it, deep \htwo\ cores cannot form and the gas is unable to cool because \htwo\ is a critical coolant of interstellar gas. Our self-shielding implementation uses entirely local methods, and distills the complex physics involved into a computationally expedient format optimized for large-scale galaxy simulations. In this paper we only model fixed-density situations with the hydrodynamics turned off. Our future work will be to run galaxies with the full suite of hydrogen and helium chemistry, radiative transfer, and hydrodynamics. 

There are several outstanding questions concerning the hydrogen content of observed galaxies that the molecular addition to \textsc{Ramses-RT} is uniquely poised to answer. Traditionally the \htwo\ content of galaxies is calculated from a conversion factor between the easily observable carbon monoxide (CO) and the elusive \htwo. However, a growing body of evidence suggests a CO-dark component to the molecular ISM \citep{Tielens1985, Wolfire2010, Smith2014}. We can explore this conversion factor by adding CO chemistry analogous to \htwo\ into \textsc{Ramses-RT} . The origins of \hi\ high-velocity clouds (HVCs) outside our galaxy and others remain a mystery \citep{Muller1963, Wakker1997, Wakker2001}. Because we now fully characterize the \hi\ content in our model, we will be able to identify HVCs and track their origin. Furthermore, we can bring our chemistry model to a cosmological context. This realm hosts the `too big to fail' problem \citep{Boylan-Kolchin2011} where Lambda cold dark matter simulations predict subhaloes that are too dense to host any measured satellites from matching observed galaxies \citep{Papastergis2014}. The ALFALFA survey \citep{Haynes2011} infers the size of such galaxies via \hi\ measurements, and our model can be a useful tool to connect these observations to cosmological galaxy simulations. By modelling the \htwo\ chemistry on a cell-by-cell basis, we build a foundation on which to explore even the largest of galactic problems.

\section*{Acknowledgements}

We thank the anonymous reviewer for their many insightful comments and suggestions. SN is supported by the University of Z\"urich Candoc Scholarship, and performed these simulations on the Piz Daint supercomputer in the Swiss National Supercomputing Centre in Lugano. JR was funded by the European Research Council under the European Union's Seventh Framework Programme (FP7/2007-2013) / ERC Grant agreement 278594-GasAroundGalaxies and the ORAGE project from the Agence Nationale de la Recherche under grant ANR-14-CE33-0016-03.


\bibliographystyle{mnras}
\bibliography{mendely,sln_refs_ads} 





\bsp	
\label{lastpage}
\end{document}